\pdfoutput=1
\documentclass[11pt]{article}

\usepackage[T1]{fontenc}
\usepackage[sc]{mathpazo}
\usepackage[margin=1in]{geometry}
\usepackage{microtype}
\usepackage{amsthm,amsmath,amsfonts,amssymb}
\usepackage[authoryear,round]{natbib}
\usepackage{graphicx}
\usepackage{booktabs}
\usepackage[font=small,labelfont=bf,labelsep=period]{caption}
\usepackage[colorlinks,citecolor=blue,urlcolor=blue,linkcolor=blue]{hyperref}
\theoremstyle{plain}
\newtheorem{proposition}{Proposition}
\theoremstyle{definition}

\title{How much can a human being lift? A Bayesian latent capability
model for Olympic weightlifting}
\author{Manuele Leonelli\\[4pt]
\normalsize School of Science and Technology, IE University, Madrid, Spain\\
\normalsize \href{mailto:manuele.leonelli@ie.edu}{manuele.leonelli@ie.edu}
\quad ORCID \href{https://orcid.org/0000-0002-2562-5192}{0000-0002-2562-5192}}
\date{}

\begin{document}

\maketitle

\begin{abstract}
Estimates of the limits of human performance are usually built on record books, on the premise that a record is the best an athlete could do. In
Olympic weightlifting the premise fails, because the competitor declares the
weight before each of three attempts per lift and only the heaviest
successful attempt is recorded, so a record is a lower bound, chosen by the
athlete, on a maximal capability never observed. We treat the
maximal lift of each competitor on each competition day as a latent
variable, of which every attempt is a trial and every declared weight a
measurement. Its distribution at a given bodyweight follows an extreme value mixture model with an estimated threshold, and an
athlete-specific random effect links the competitions of each athlete. The model is fitted to the $117{,}892$ attempts of $5{,}040$
athletes in $141$ international competitions between 2017 and 2026. The
upper tail of capability is bounded, with posterior probability above $0.999$ for
men, $0.96$ for the women's snatch and $0.94$ for the women's clean and jerk. At
$110$ kilograms of bodyweight the men's ceiling lies one to three per cent
above the heaviest lifts, against twelve and sixteen per cent for women
at $86$ kilograms. The record book alone places the
men's ceiling six and twelve kilograms higher, because the best lift of a day ending on a success overstates capability.


\end{abstract}

\section{Introduction}
\label{sec:intro}

Consider a lifter who opens the snatch at $180$ kilograms and makes it,
raises to $185$ and makes it, and then misses at $190$. The official result
is $185$ kilograms. It shows that the lifter's capability on that day was at
least $185$ kilograms, but not by how much it exceeded that value. The miss
at $190$ is weak evidence of an upper limit, since a lift can fail through
timing, bar position or nerve as well as through lack of strength, and
$188$ kilograms might have succeeded. The opening weight of $180$ kilograms
is stronger evidence about where the lifter placed that limit, because
lifters choose an opener they expect to lift. The recorded result is thus one of three
choices made by an athlete who knows more about his or her own capability
than the result reveals.

Statistical estimates of the limits of human performance start from record
books. \citet{einmahl2008} treated the best performances in athletics as
draws from a distribution and estimated its right endpoint, the limit of the
discipline, from the observed extremes. The same approach has since been
applied to sprinting \citep{einmahl2011,einmahl2026}, to longevity
\citep{aarssen1994,alves2017}, to swimming, with progressively richer models
\citep{spearing2021,spearing2023}. In every case the recorded result is taken
to be the performance. In Olympic weightlifting that assumption fails, and
it fails in a structured way. A competitor has three attempts at the snatch
and three at the clean and jerk. The weight is declared before each attempt,
may never decrease, and the heaviest success is credited. About half of all
attempt sequences in international competition contain both a made lift and
a heavier miss, so the capability of those competitors is bracketed from
both sides. Most of the rest end on a successful final attempt, so the
capability of those competitors is bounded only from below. The second group
contains the champions, whose lifts matter most for the upper tail and about
whom the record says least.

A natural starting point is a latent variable model in which each
competitor has an unobserved capability on each day and each attempt is a
pass-or-fail trial whose success probability falls as the bar rises above
that capability. The difficulty lies in how the weights are chosen. They
are not fixed in advance but set in sequence by athletes who know their own
strength, in response to the outcomes of earlier attempts. After a
successful opener the second attempt succeeds about six
times in ten whether the bar is raised by two kilograms or by ten, so the
declared weights carry much of the information about capability and the
outcomes little. A model that treats the weights as a fixed
design ignores that information.

This paper models the unobserved capability together with the choices made
about it. Each competitor on each competition day is given a true maximal
lift for the snatch and one for the clean and jerk. Across competitors of
a given bodyweight these capabilities follow an extreme value mixture
model \citep{behrens2004,scarrott2012}, with a mixture of
normal distributions for the body and a generalized Pareto tail above a
threshold that is estimated with everything else. The shape of that tail decides whether strength is bounded. Every attempt is a trial against
the capability, with a logistic success probability whose scale is the
amount of luck in a lift. Every declared weight is a measurement of it,
set below capability by a margin whose typical size and spread depend on
the situation in which the weight was chosen, an opener, a raise after a
success or a raise after a miss, subject to the rule that the bar must go
up after a success.

Not everything can be learned from competition data. If every declared
weight were placed further below capability and the attempt-order effects
adjusted accordingly, the likelihood would change only through the outcomes
of first attempts. The data therefore say little about how far below their limit athletes declare. This margin decides the distance between a record and the capability behind it. Left free, the fit moves towards athletes who declare far below
their limit and attempts whose outcome is close to a coin flip, a
description that the measured reliability of maximal strength rules out. We
therefore fix the scale of attempt-to-attempt variation from that external
evidence and report how the results change with it.

The upper tail of capability is bounded with posterior probability above $0.999$ for men. For the women's two lifts it is $0.96$ and $0.94$. At $110$
kilograms of bodyweight the men's ceiling lies one
to three per cent above the heaviest lifts since 2019. The women's ceilings at $86$ kilograms lie twelve and sixteen per cent above theirs.
Fitted to the best lifts alone, the same
distribution places the men's ceiling six and twelve kilograms higher,
because the best lift of a day ending on a successful attempt overstates
capability.

The capability distribution also has a practical use. To compare lifters
of different bodyweights, the sport multiplies each total by a coefficient
that depends on bodyweight \citep{sinclair1985}, refitted to the world
records every Olympic cycle. A lift could instead be scored by its position
in the capability distribution at the lifter's bodyweight, such as the
quantile it reaches, which does not depend on a handful of records. More
generally, whenever performances come from
challenges that athletes choose for themselves, those choices carry
information about ability and belong in the model, although how far below
their limit athletes choose to go cannot be learned from competition data
alone.

\section{Data}
\label{sec:data}

The data are the results of the senior international competitions
sanctioned by the sport's world governing body between March 2017 and May
2026, namely world and continental championships, Olympic Games, World Cup
and Grand Prix meetings, and the qualifying events of each Olympic cycle. They
come from the open OpenWeightlifting project
(\url{https://github.com/euanwm/OpenWeightlifting}), which compiles the governing body's result sheets. The processed attempt-level data and all code are available at
\url{https://github.com/manueleleonelli/weightlifting-capability}. For
every competitor at every competition the archive records name, bodyweight at the weigh-in to two decimals, the weight on the bar at each of six attempts and whether each attempt succeeded.

The unit of analysis is the athlete-day-lift, one competitor, one
competition and one of the two disciplines, with up to three attempts. The
data contain $117{,}892$ attempts by $5{,}040$
athletes, $2{,}646$ men and $2{,}394$ women, in $141$ competitions. They
make up $39{,}883$ athlete-day-lifts on $20{,}198$ athlete-days, of which
$19{,}685$ have both lifts. The same athlete appears in four competitions
on average and in up to thirty-seven (Section~\ref{sec:method-athlete}).

The archive goes back to 1998 (Table~\ref{tab:cov}), but the earlier years
differ from the present in two ways that a smooth time trend cannot absorb.
The weight categories of that era were abolished in 2018 together with every standing world record. The re-analysis of stored samples from the
2008 and 2012 Olympic Games produced sixty-one retrospective doping
violations in weightlifting, thirty-four of them by medallists
\citep{kolliari2021}. At $86$ kilograms of bodyweight, for instance, the
heaviest women's lifts of all years, from 2009 and 2012, have not
been approached since. A model fitted to all years would carry such lifts
to the present through its time trend. The analysis therefore starts in
March 2017, and Section~\ref{sec:s-archive} of the supplement
repeats it
on the whole archive.

\begin{table}[t]
\centering
\caption{Coverage of the archive by period.}
\label{tab:cov}
\begin{tabular}{lrrr}
\toprule
Period & Competitions & Athlete-day-lifts & Athletes \\
\midrule
1998--2004 & 14 & 5{,}865 & 1{,}602 \\
2005--2011 & 50 & 14{,}379 & 2{,}958 \\
2012--2016 & 69 & 18{,}259 & 3{,}475 \\
2017--2018 & 33 & 9{,}012 & 2{,}167 \\
2019--2026 & 108 & 30{,}871 & 4{,}055 \\
\bottomrule
\end{tabular}
\end{table}

Bodyweight enters the model as a continuous covariate, not through the weight categories, whose boundaries were redrawn in 1993, 1998, 2018 and
2025, twice within the analysis period. A record series by category breaks at each such revision, unlike bodyweight.

\section{Background}
\label{sec:background}

\subsection{Extreme value models for sporting records}

Three lines of work bear on the problem. The first estimates the right
endpoint of a performance distribution directly from the top performances.
\citet{einmahl2008} did this for athletics, obtaining both an ultimate record for each event and a
measure of how close the current record is to it. \citet{einmahl2011}
applied the method to the 100 metres, and \citet{einmahl2026} refined it,
finding the men's record within hundredths of a second of its estimated limit, with a much larger margin for women.
The same tools have asked whether human longevity is bounded
\citep{aarssen1994,alves2017} and how far the dominance of three players in
men's tennis lay beyond the rest of the field \citep{leonelli2026exceptional}.

The second line models the whole population of performances rather than
only the best. \citet{stephenson2013} built a model for athletics in which
all performances are described at once, an idea anticipated by the record
progression analysis of \citet{berthelot2008}. \citet{robinson1995} asked
whether an individual record, the women's 3000 metres world record set in
1993, is anomalous relative to the distribution that produced it. \citet{nguyen2026} describe
the ten events of the decathlon with a Bayesian compositional model with
athlete effects and find a score above $9{,}200$ points possible but
unlikely.

The third line, closest to the present work, is \citet{spearing2021}, who model swim times through their rate of occurrence with covariates in location and scale. They were motivated by a defect of the official points system, which scores a swim against the current world record, itself a random extreme, rather than against the whole distribution. Weightlifting's Sinclair coefficient \citep{sinclair1985}, recalibrated on records each Olympic cycle, has the same defect.

\subsection{Latent trials, adaptive designs and informative censoring}

A made lift bounds capability from below and a missed lift bounds it,
weakly, from above, so the natural description of the problem is as
censoring. The literature on extremes under censoring
\citep{beirlant2007,einmahl2008censoring,worms2014} assumes censoring
independent of the quantity of interest. Here it is not. The athlete
chooses the weights knowing his or her capability, so the
censoring is informative \citep{little2019}.

A pass-or-fail trial whose success probability depends on the difference
between a latent ability and a known difficulty is the Rasch model of item
response theory \citep{rasch1960}. Two of its properties matter here. The
number of successes is sufficient for the ability, so the conditional
likelihood given that number involves the difficulties alone
\citep{andersen1970}, a device that has been proposed for identifying the
scale of the trial without a model for the ability. In adaptive testing the
items are chosen in sequence in the light of earlier responses, and the
likelihood that conditions on the chosen items remains valid when the
selection depends on the observed responses alone, but not when it depends
on the ability itself \citep{mislevy2000}. The same distinction is Rubin's
condition for an ignorable design \citep{rubin1976}. The declared weights
of a weightlifter are chosen in the light of the ability, so neither the
conditional likelihood nor the likelihood that conditions on the weights is
valid, and Section~\ref{sec:method} replaces conditioning with a model for
the choice.

\subsection{Extreme value mixture models and dependence}

Any model with a generalized Pareto tail has to say where the tail begins.
The classical approach fixes a threshold from diagnostic plots and fits the
exceedances alone \citep{davison1990,coles2001}. Extreme value mixture
models \citep{scarrott2012} instead combine a model for the bulk below the
threshold with a generalized Pareto distribution above it and estimate the
threshold with everything else. \citet{behrens2004} introduced the
construction with a parametric bulk, and \citet{nascimento2012} made the
bulk a mixture of gamma distributions. This paper uses a mixture of normal
distributions instead, since the standardized error of capability takes
values on the whole real line, and estimates the threshold inside a latent
variable model.

For the dependence between two extremes the basic distinction is between
asymptotic dependence, where two variables that are both extreme remain
related however far out one looks, and asymptotic independence, where the
relation fades \citep{ledford1996,coles1999}. The conditional model of
\citet{heffernan2004} and the family of \citet{wadsworth2017} fit data without deciding the class in advance. Bivariate analysis of Olympic results goes back at least to \citet{adam2012}.

\subsection{Statistical work on weightlifting}

The literature specific to the sport is mostly about scaling performance by
bodyweight, from the dimensional argument of \citet{lietzke1956} and the
record analysis of \citet{croucher1984} to the coefficient of
\citet{sinclair1985} that is still in official use and later revisions
\citep{ford2000}. Bayesian hierarchical models of individual performance
are common in sports statistics \citep{jensen2009}. The closest modelling work is \citet{griffin2022}, who build Bayesian models of
how individual sprinters and weightlifters progress with age, extended to
within-season trajectories by \citet{griffin2026}. That work models the body
of the performance distribution and takes the recorded best as the
performance. Attempt selection has been studied descriptively, mostly in
powerlifting \citep{travis2021}, and responds to competitive position,
which is why the censoring cannot be ignored. Neither the tail of
strength nor the attempt structure has been modelled before.

\section{Methodology}
\label{sec:method}

The model has four layers. The first is the distribution of capability
across athlete-days at a given bodyweight and year, whose upper endpoint is
the ultimate limit. The second relates capability to the success of each
attempt, and the third treats each declared weight as a measurement of
capability. The fourth links the competitions of the same athlete, and
Section~\ref{sec:method-lik} combines the layers into a likelihood.

\subsection{The distribution of capability}
\label{sec:method-marginal}

Index by $i$ an athlete-day, one competitor at one competition, and by
$l \in \{s, c\}$ the two disciplines, snatch and clean and jerk. Let $Z_i^{l}$
be the heaviest weight that competitor $i$ could have raised in discipline
$l$ on that day. We call it the capability, and Section~\ref{sec:method-latent}
makes the definition exact. Capability depends on the size of the athlete and
on the era, and is modelled as
\begin{equation}
\label{eq:capability}
Z_i^{l} = \mu_l(x_i) + \omega_l(x_i)\, \varepsilon_i^{l},
\end{equation}
a location and a scale that shift with covariates, and a standardized error
that carries the shape of the distribution. The covariates are three. The
first is bodyweight at the weigh-in, entered as
$b_i = (\log \text{bodyweight}_i - \bar{m})/s$, centred and scaled by the
sample mean and standard deviation of log bodyweight. The second is its
square, made orthogonal to the intercept and to $b_i$, scaled to unit standard deviation and written $b_i^{(2)}$. The third is the competition year
centred on its sample mean, $t_i$. The location is linear in all three,
\begin{equation}
\label{eq:mu}
\mu_l(x_i) = \beta_{l0} + \beta_{l1}\, b_i + \beta_{l2}\, b_i^{(2)} +
\beta_{l3}\, t_i,
\end{equation}
so that capability rises with bodyweight at a rate that is allowed to flatten
for the heaviest athletes, and drifts with time. The scale is log-quadratic
in bodyweight,
\begin{equation}
\label{eq:omega}
\log \omega_l(x_i) = \gamma_{l0} + \gamma_{l1}\, b_i + \gamma_{l2}\, b_i^{(2)},
\end{equation}
so that the spread of capability among athletes of the same size may widen
or narrow with size, and differently at the two ends of the range. Equation~\eqref{eq:capability} is a regression, in
which $\mu_l(x_i)$ locates the capability of athletes of that size and year while $\omega_l(x_i)$ measures their spread.

The error $\varepsilon_i^{l}$ carries the shape of the distribution. The
population of athlete-days is drawn from fields of very different depth,
from continental championships and regional games, whose competitors trail
far below the typical lift, to world championships and Olympic Games, whose
competitors are compressed into a narrow band above it. The body of the
error is therefore a mixture of three normal distributions,
\begin{equation}
\label{eq:bulk}
h_l(e) = \sum_{k=1}^{3} w_{lk}\, \frac{1}{s_{lk}}\,
\phi\!\left( \frac{e - m_{lk}}{s_{lk}} \right),
\qquad
H_l(e) = \sum_{k=1}^{3} w_{lk}\, \Phi\!\left( \frac{e - m_{lk}}{s_{lk}} \right),
\end{equation}
where $\phi$ and $\Phi$ are the standard normal density and distribution function. The weights $w_{lk}$ are positive with unit sum. The components have means $m_{lk}$ and standard deviations $s_{lk}$. The first component has $m_{l1} = 0$ and $s_{l1} = 1$, which fixes the location and scale of the error. Then $\mu_l(x)$ is the centre of the reference component and $\omega_l(x)$ its spread in kilograms. The other two
components are free, and with them the body can be skewed, heavy on either
side or bimodal as the fields require. The upper tail is where extreme value
theory enters. Above a threshold $u_l$ the error follows a generalized
Pareto distribution,
\begin{equation}
\label{eq:splice}
g_l(e) =
\begin{cases}
h_l(e), & e \le u_l, \\[4pt]
\{1 - H_l(u_l)\}\, \dfrac{1}{\sigma_l}
\left( 1 + \xi_l\, \dfrac{e - u_l}{\sigma_l} \right)^{-1/\xi_l - 1},
& e > u_l,
\end{cases}
\end{equation}
where $\xi_l$ is the shape of the tail and $\sigma_l > 0$ its scale. The tail carries the mass $1 - H_l(u_l)$ that the body places above the threshold, so that the two pieces integrate to one. This is the extreme
value mixture model of \citet{behrens2004} with the mixture body of
\citet{nascimento2012}, with normal components in place of gamma components
because the error takes values on the whole real line. The density may jump
at the threshold, as in those papers, since nothing in the problem requires
the body and the tail to meet smoothly. The generalized Pareto is the form
that exceedances of a high threshold take for any distribution with a
well-behaved tail \citep{davison1990,coles2001}, and its shape answers the
question of the paper. For $\xi_l < 0$ the distribution has a finite upper endpoint. For $\xi_l = 0$ the tail is exponential, for $\xi_l > 0$ heavier than exponential, and in both cases unbounded. The regression in
\eqref{eq:capability} removes location and scale before the tail is
modelled, as in \citet{mcneil2000}, so that the standardized error is the
athlete's position relative to competitors of the same size and era.

Write $G_l$ for the distribution function of \eqref{eq:splice}, equal to
$H_l(e)$ for $e \le u_l$ and to $1 - \{1 - H_l(u_l)\}\{1 + \xi_l (e -
u_l)/\sigma_l\}^{-1/\xi_l}$ above the threshold. Two quantities follow. For
a probability $p > H_l(u_l)$, the capability exceeded by a fraction $1 - p$
of athlete-days at covariate value $x$ is
\begin{equation}
\label{eq:quantile}
Q_l(p; x) = \mu_l(x) + \omega_l(x) \left[ u_l +
\frac{\sigma_l}{\xi_l}
\left\{ \left( \frac{1 - p}{1 - H_l(u_l)} \right)^{-\xi_l} - 1 \right\}
\right],
\end{equation}
and when $\xi_l < 0$ the distribution has the finite upper endpoint
\begin{equation}
\label{eq:endpoint}
\zeta_l(x) = \mu_l(x) + \omega_l(x)
\left( u_l + \frac{\sigma_l}{-\xi_l} \right).
\end{equation}
Equation~\eqref{eq:endpoint} is what an ultimate limit means in this paper,
the heaviest weight that any competitor of bodyweight and era $x$ could
raise. Uncertainty about it has two parts, the probability $P(\xi_l \ge 0)$
that no such limit exists and the distribution of $\zeta_l(x)$ given that
one does. Because $\zeta_l \to \infty$ as $\xi_l \to 0^-$, that
distribution has a heavy upper tail whenever shapes close to zero are plausible. We therefore report the probability that a ceiling exists, the probability that it lies within ten, twenty-five or fifty per cent of what has been lifted, and its median.

\subsection{From capability to attempts}
\label{sec:method-latent}

Capability is never observed. What is observed, for each athlete-day and
discipline, is a sequence of up to three declared weights
$w_{i1}^{l} \le w_{i2}^{l} \le w_{i3}^{l}$, nondecreasing by rule, with a success indicator $S_{ij}^{l}$ for each. The attempt model says how the
outcomes arise from $Z_i^{l}$. Whether an attempt succeeds depends on more
than strength, namely on the timing of the pull, the position of the catch
and the athlete's nerve at that moment. We represent this by giving each attempt its
own momentary maximum, $Z_i^{l} \exp\{\tau_l (\delta_j^{l} + L_{ij}^{l})\}$,
where $L_{ij}^{l}$ is a standard logistic variable independent across
attempts, and the attempt succeeds when the declared weight lies below that
moment's maximum. The spread of the momentary maximum is proportional to
capability, because a lifter whose maximum is $200$ kilograms varies from
attempt to attempt by more kilograms than one whose maximum is $100$, while the
relative scale $\tau_l$ is the same for both. Later attempts are allowed to
be systematically easier or harder than the first at the same margin
through $\delta_j^{l}$, with $\delta_1^{l} = 0$. Since $L$ has distribution
function $F(v) = (1 + e^{-v})^{-1}$ and is symmetric about zero,
\begin{equation}
\label{eq:link}
P(S_{ij}^{l} = 1 \mid Z_i^{l}) =
F\!\left( \frac{\log(Z_i^{l} / w_{ij}^{l})}{\tau_l} + \delta_j^{l} \right).
\end{equation}
Equation~\eqref{eq:link} is what defines capability. A first attempt at
exactly $Z_i^{l}$ succeeds with probability one half, an attempt a fraction
$\tau_l$ below it succeeds with probability $0.73$, and one $2\tau_l$ below
with probability $0.88$. Conditionally on $Z_i^{l}$ and the shifts, the
three attempts are independent. With $\log Z_i^{l} / \tau_l$ as the ability
and $\log w_{ij}^{l} / \tau_l - \delta_j^{l}$ as the difficulty,
\eqref{eq:link} is a Rasch model \citep{rasch1960} with continuous, known
difficulties.

\subsection{Declared weights as measurements}
\label{sec:method-decl}

A declared weight is the athlete's own statement about his or her
capability, which we treat as a measurement with an error that depends on
the situation in which the weight was chosen. Combined with the attempts of
Section~\ref{sec:method-latent}, this gives a selection model in the sense
of \citet{little2019}. There are three situations, $k = 1, 2, 3$. The opener
is chosen before any outcome is known. A raise after a successful attempt
must exceed the previous bar by rule. A raise after a missed attempt is the
minority choice, since about three athletes in four repeat the weight after a miss. A repeat declares nothing new and enters the model only through its outcome. In situation $k$ the declared
weight at attempt $j$ is capability reduced by a margin,
\begin{equation}
\label{eq:margin}
\log w_{ij}^{l} = \log Z_i^{l} - R_{ij}^{l},
\end{equation}
whose distribution, as for a measurement error, is specified given the quantity measured. Given $Z_i^{l}$, the margin has a two-piece normal
distribution with mode $m_k^{l}$, the margin the athlete typically leaves.
Its scale $s_k^{l,+}$ above the mode describes how far below capability
some athletes declare, and its scale $s_k^{l,-}$ below the mode how far
above capability they reach. Both raises, $k = 2, 3$, add at least one kilogram to the bar, $w_{ij}^{l} \ge w_{i,j-1}^{l} + 1$, which truncates the margin from above at $\log\{Z_i^{l} / (w_{i,j-1}^{l} + 1)\}$. The density is divided by the two-piece normal distribution function there. Otherwise the margin is independent of capability, so the declarations follow the distribution of capability, shifted down and spread by
the margins.

\subsection{Repeated competitions of the same athlete}
\label{sec:method-athlete}

The population of athlete-days is not exchangeable. The analysis set
contains $2{,}646$ men and $2{,}394$ women, who appear in up to thirty-seven competitions each. The top of the capability distribution is made up of the repeated competitions of a few champions. We let the competitions of the
same athlete share a factor on the probability scale. Writing
$U_i^{l} = G_l(\varepsilon_i^{l})$ for the error on the uniform scale and
$C_i^{l} = \Phi^{-1}(U_i^{l})$ for the same error on the normal scale, the
competitions of athlete $a$ have
\begin{equation}
\label{eq:athlete}
C_i^{l} = \sqrt{\rho_{a,l}}\, V_a^{l} + \sqrt{1 - \rho_{a,l}}\, \epsilon_i^{l},
\end{equation}
with $V_a^{l}$ and $\epsilon_i^{l}$ independent standard normal. Every
athlete-day keeps the marginal distribution \eqref{eq:splice}, so its quantiles and endpoint are
unchanged, while the correlation $\rho_{a,l}$ between two competitions of
the same athlete measures how much of the spread of capability is
persistent.

\subsection{The two disciplines together}
\label{sec:method-dep}

A competitor's snatch and clean and jerk on the same day are related, since they
draw on the same body, the same preparation and the same form. The relation
is placed on the errors on the uniform scale of
Section~\ref{sec:method-athlete}. The pair
$(U_i^{s}, U_i^{c})$ is given a bivariate t copula with correlation $\rho$
and degrees of freedom $1/\psi$, $\psi \in [0, 1/2]$. The correlation says
how strongly the two disciplines move together in the body of the
distribution. The degrees of freedom say whether they move together in the
extremes. For $\psi = 0$ the copula is Gaussian and the two disciplines are
asymptotically independent \citep{ledford1996}. The probability that the
clean and jerk is extreme given that the snatch is extreme then goes to zero as
the level rises. For any $\psi > 0$ the copula is asymptotically dependent
and that probability has a positive limit. Which of the two holds decides
whether a competitor can be at the top of both lifts on the same day, and so how large a total can be.

\subsection{The likelihood}
\label{sec:method-lik}

For athlete-day-lift $(i, l)$ write $\eta_{ij}^{l}(z) = \log(z /
w_{ij}^{l}) / \tau_l + \delta_j^{l}$ for the argument of the logistic
distribution function $F$ in \eqref{eq:link} at capability $z$, $\mathcal{J}_i^{l}$ for the set of attempts whose weight is
a declaration in one of the three situations of
Section~\ref{sec:method-decl}, $k_{ij}$ for the situation of attempt
$j \in \mathcal{J}_i^{l}$, and $d_k^{l}$ and $D_k^{l}$ for the two-piece
normal density and distribution function of the margin in situation $k$. A
weight repeated after a miss is left out of $\mathcal{J}_i^{l}$. Given capability $z$, the attempts and declarations of
the athlete-day-lift contribute
\begin{equation}
\label{eq:unitlik}
\ell_i^{l}(z) = \prod_{j} F\{\eta_{ij}^{l}(z)\}^{S_{ij}^{l}}
\left[1 - F\{\eta_{ij}^{l}(z)\}\right]^{1 - S_{ij}^{l}}
\prod_{j \in \mathcal{J}_i^{l}}
\frac{d_{k_{ij}}^{l}\{\log(z / w_{ij}^{l})\}}
{w_{ij}^{l}\, D_{k_{ij}}^{l}[\log\{z / (w_{i,j-1}^{l} + 1)\}]},
\end{equation}
where the first product runs over all attempts taken and $D_1^{l} \equiv 1$,
because the opener is not truncated. By \eqref{eq:capability} and
\eqref{eq:splice} the capability of athlete-day $i$ has density and
distribution function
\[
g_i^{l}(z) = \frac{1}{\omega_l(x_i)}\, g_l\!\left\{ \frac{z -
\mu_l(x_i)}{\omega_l(x_i)} \right\},
\qquad
G_i^{l}(z) = G_l\!\left\{ \frac{z - \mu_l(x_i)}{\omega_l(x_i)} \right\}.
\]
Given the athlete factor $V_a^{l} = v$, the athlete-days of athlete $a$ are
independent, and by \eqref{eq:athlete} the conditional density of $Z_i^{l}$
is $g_i^{l}(z)\, \kappa_{\rho_{a,l}}[\Phi^{-1}\{G_i^{l}(z)\}, v]$, where
\[
\kappa_\rho(c, v) = \frac{1}{\sqrt{1 - \rho}\, \phi(c)}\,
\phi\!\left( \frac{c - \sqrt{\rho}\, v}{\sqrt{1 - \rho}} \right)
\]
is the ratio of the conditional to the marginal density of $C_i^{l}$. The
likelihood of discipline $l$, with parameters $\theta_l$, is therefore
\begin{equation}
\label{eq:lik}
L_l(\theta_l) = \prod_{a} \int_{-\infty}^{\infty} \prod_{i \in a}
\left( \int_0^{\infty} \ell_i^{l}(z)\, g_i^{l}(z)\,
\kappa_{\rho_{a,l}}[\Phi^{-1}\{G_i^{l}(z)\}, v]\, dz \right) \phi(v)\, dv.
\end{equation}
For each athlete, the capability in every competition is integrated out
given the athlete factor, which is then integrated out. With
$\rho_{a,l} = 0$, $\kappa = 1$ and the athlete-days are independent.

The two disciplines are joined in a second stage
(Section~\ref{sec:method-comp}), in which the athlete factor is left out
and the athlete-days are treated as independent. A competitor who completes
both lifts on day $i$ contributes an integral over the two capabilities,
\begin{equation}
\label{eq:likcop}
L_{sc}(\rho, \psi) = \prod_{i} \iint \ell_i^{s}(z^{s})\, \ell_i^{c}(z^{c})\,
g_i^{s}(z^{s})\, g_i^{c}(z^{c})\,
c_{\rho, \psi}\{G_i^{s}(z^{s}), G_i^{c}(z^{c})\}\, dz^{s}\, dz^{c},
\end{equation}
where $c_{\rho, \psi}$ is the density of the t copula of
Section~\ref{sec:method-dep}.

\subsection{Identification}
\label{sec:method-ident}

Comparing declarations with outcomes in \eqref{eq:lik} might seem to identify $\tau_l$, since the declarations locate capability relative to the bar and the outcomes show how often a bar at that distance is lifted. Proposition~\ref{prop:ridge} shows that this
holds only in part.

\begin{proposition}
\label{prop:ridge}
Fix $\tau_l$ and let $\theta_l$ collect the other parameters of discipline
$l$. For $\Delta \in \mathbb{R}$, let $T_\Delta \theta_l$ be obtained from
$\theta_l$ by multiplying the location coefficients $\beta_{l0}, \dots,
\beta_{l3}$ by $e^{\Delta}$, adding $\Delta$ to the scale intercept
$\gamma_{l0}$ and to the margin modes $m_1^{l}, m_2^{l}, m_3^{l}$, and
subtracting $\Delta / \tau_l$ from the attempt-order shifts $\delta_2^{l}$
and $\delta_3^{l}$, with all other parameters unchanged. Then, for every
$\theta_l$ and $\Delta$,
\[
L_l(T_\Delta \theta_l) = L_l^{(\Delta)}(\theta_l),
\]
where $L_l^{(\Delta)}$ is the likelihood \eqref{eq:lik} with
$\eta_{i1}^{l}(z)$ replaced by $\eta_{i1}^{l}(z) + \Delta / \tau_l$ in
\eqref{eq:unitlik} for every athlete-day-lift. In particular, the
likelihood without the outcomes of the first attempts is constant along
$\{T_\Delta \theta_l : \Delta \in \mathbb{R}\}$.
\end{proposition}

The proof is in Section~\ref{sec:s-proof} of the supplement. The level of capability relative to the declared weights, which decides how
far a record lies below the ceiling, is therefore learned only from the
outcomes of the openers, and only for a given $\tau_l$. A larger $\tau_l$
with all margins larger can leave the opener success rate unchanged
while weakening the dependence of each outcome on its margin. Since $\tau_l$ and $\Delta$ move together, the data determine them weakly. Left
free, the fit drifts along the direction of Proposition~\ref{prop:ridge} to
a scale of $14$ per cent (Section~\ref{sec:results-tau}).

The comparison with a Rasch model shows why. A Rasch model sets the scale
of the logistic to one by convention, which defines the unit of ability,
and with known difficulties it identifies the level of ability as well.
Here the unit is fixed by the kilogram. The scale $\tau_l$ is the variability of a lifter from one attempt to the next, so its value changes the distance between a record and the ceiling. The difficulties are known but chosen by
the athletes a few per cent below their capability, through margins whose
mode is unknown, and the attempt-order shifts are free, which leaves only
the openers to fix the level. Attempts far from capability, which would
reveal the slope of the success curve, are almost never made.

Physiology bounds $\tau_l$. The test-retest coefficient of variation of a
one-repetition maximum has a median of $3.3$ per cent among participants
with resistance-training experience in the studies reviewed by
\citet{grgic2020}, and the variation
between attempts minutes apart on the same day cannot exceed the variation
between sessions. As \citet{manski2003} and \citet{gustafson2015} recommend
for parameters that the data do not pin down, we bring in this outside knowledge, reporting the conclusions as a function of it. We fix $\tau_l$ at $0.02$, for which the momentary maximum has a
standard deviation of $3.6$ per cent of capability and a lift two per cent
under capability succeeds three times in four. Every quantity of interest is also reported at $0.01$ and $0.03$. The bound implies
$\tau_l \le 0.018$, so the reference value lies slightly above it. It is
kept as a round value between an anchor well inside the bound and one outside it. The conclusions change little across that range
(Section~\ref{sec:results-pred}).

\subsection{Priors}
\label{sec:method-prior}

Inference is Bayesian. Every parameter except the attempt scale $\tau_l$,
fixed in Section~\ref{sec:method-ident}, receives a prior distribution. All
statements about uncertainty in the paper are posterior probabilities and intervals. The posterior is explored by Markov chain Monte Carlo
(Section~\ref{sec:method-comp}). The
priors are weakly informative. The location coefficients have
$\beta_{lk} \sim N(0, 300^2)$. The scale coefficients have $\gamma_{l0}
\sim N(\log 15, 0.7^2)$, so that the spread of capability among athletes of
average size is centred on $15$ kilograms, and $\gamma_{l1}, \gamma_{l2}
\sim N(0, 1)$.

For the body, the means of the two free components have $m_{l2}, m_{l3}
\sim N(0, 3^2)$, the log standard deviations $\log s_{lk} \sim N(0, 1)$ on
$(0.05, 20)$, and the weights are $w_{lk} \propto e^{a_{lk}}$ with
$a_{l1} = 0$ and $a_{l2}, a_{l3} \sim N(0, 1.5^2)$. The two free components
have the same prior and are not ordered, because an ordering creates a
boundary at which the sampler sticks when they overlap. Every quantity reported below is invariant to their labels.
The threshold is a parameter, with $u_l \sim N(1.5, 1)$ on $(0.2, 4)$,
centred between the $0.9$ and $0.95$ quantiles of the reference component,
so that the uncertainty about where the tail begins is carried into every
quantity derived from the tail. The tail has $\log \sigma_l \sim N(\log 0.5,
1)$ on $(0.01, 10)$ and $\xi_l \sim N(0, 0.3^2)$ truncated to $(-1.5, 0.4)$, which gives
probability $0.55$ to a bounded tail. Reweighting the posterior draws to a
wider prior for $\xi_l$ or $u_l$, or to a threshold below $3$, moves the
median ceilings by at most $1.1$ kilograms (Section~\ref{sec:s-sens} of the supplement).
The wider prior for the threshold lowers the probability of a finite
ceiling for the women's clean and jerk from $0.94$ to $0.89$ and moves the
other three by at most $0.01$.

The attempt-order shifts have $\delta_j^{l} \sim N(0, 1.5^2)$, the margin
modes $m_k^{l} \sim N(0.03, 0.1^2)$ on $(-0.2, 0.3)$ and the log scales of the margins
$N(\log 0.03, 1)$ on $(0.003, 1)$. The athlete correlation $\rho_{a,l}$ is uniform on $(0, 0.95)$. The copula parameter $\rho$ is uniform on $(0.88, 0.98)$ and $\psi$ on $[0, 0.5]$, ranges whose upper bounds the posterior does not reach.

\subsection{Estimation}
\label{sec:method-comp}

The capabilities are integrated out of \eqref{eq:lik} by quadrature rather
than sampled. The inner integrand is smooth except at the threshold and at
the modes of the margin densities, some of whose scales are as small as
$0.3$ per cent of the bar, so the range is cut at these points, each piece integrated by Gauss--Legendre quadrature. The athlete factor
is integrated by twelve-point Gauss--Hermite quadrature, centred and scaled for each athlete. Both rules agree with brute-force integration (Section~\ref{sec:s-quad} of the supplement). The likelihood of one
discipline has twenty-eight parameters and takes about a fifth of a second
to evaluate. Integrating the capabilities out also helps with the
threshold. When a threshold is estimated from observed data, the likelihood
jumps as observations cross it. Here $u_l$ enters only through the integrand,
and the likelihood is smooth in it.

The posterior is harder to explore. Raising the threshold moves mass from
the body to the tail, which the scale of the generalized Pareto, the free
mixture components and the regression coefficients compensate, so these
parameters lie along curved ridges. Two devices address the problem. The
first replaces $\sigma_l$ in the sampler by the threshold-stable scale
$\sigma_l^{*} = \sigma_l - \xi_l u_l$ \citep[Section 4.3.4]{coles2001},
which does not change when the threshold moves and in terms of which the
endpoint of the standardized error is $\sigma_l^{*}/(-\xi_l)$. The second is
the sampler, differential-evolution Markov chains with sampling from the
past and snooker updates \citep{terbraak2008}, whose proposals move along differences between past states of the chains, following the shape of the posterior, curved ridges included.

Four chains, started around the posterior mode, run for $120{,}000$
generations each, or $360{,}000$ for the two women's lifts, whose posteriors
put part of their mass on shapes at or above zero. The first $30{,}000$ generations are discarded and every twentieth state of the rest is kept. The
posterior of the women's snatch has two regions, one with a higher threshold
and a shape closer to zero. They agree on the one-in-a-thousand level but
differ on the endpoint, and the reported posterior mixes them with weights
estimated by a further set of chains that moves between them. The
one-in-a-thousand level and the endpoint have $\widehat{R} \le 1.05$ in every
fit, within each region for the women's snatch (Section~\ref{sec:s-conv} of the
supplement).

The two disciplines are fitted separately in this way, and the copula in a
second stage. The likelihood \eqref{eq:likcop} is evaluated on a grid of
$(\rho, \psi)$ values at five posterior draws of the marginal parameters,
and the posterior over the grid is averaged across those draws
(Section~\ref{sec:s-copula} of the supplement). This stage
also gives draws of the pair of capabilities for every athlete-day, which
drive the dependence diagnostics of Section~\ref{sec:results-dep}.

\section{Results}
\label{sec:results}

\subsection{What the raw attempts show}
\label{sec:results-explore}

Success becomes rarer with each attempt. In the men's snatch the
proportion of successful attempts is $0.771$ at the first attempt, $0.633$
at the second and $0.437$ at the third, in the clean and jerk $0.795$, $0.601$ and $0.348$. The women's rates are within four points of the men's. The median raise is three to four per cent of the bar.

Most athlete-day-lifts do not end cleanly. In $52$ per cent of men's
snatches and $58$ per cent of clean and jerks the sequence contains both a
made lift and a heavier miss, so capability is bracketed on both sides. In
$44$ and $37$ per cent the final attempt succeeded, so capability is
bounded only from below. In $4$ and $5$ per cent every attempt failed. The
women's shares are similar. An analysis that equates capability with the best made lift
ignores the heavier miss when capability is bracketed. When the final
attempt succeeded, it treats a lower bound as the capability itself, and
this is how most of the heaviest lifts end.

\begin{figure}[t]
\centering
\includegraphics[width=0.95\textwidth]{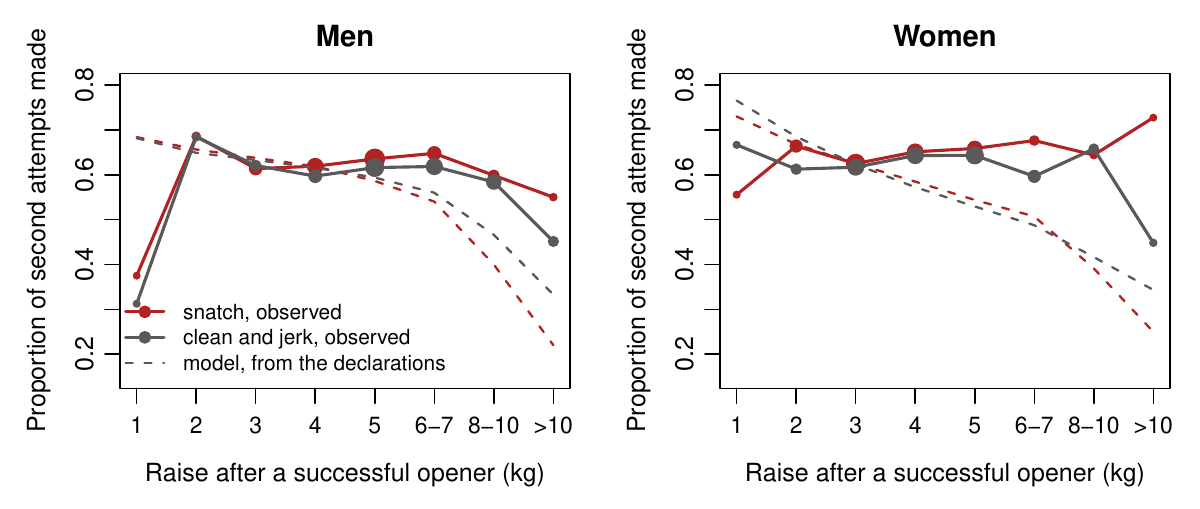}
\caption{Proportion of second attempts that succeeded after a successful
opener, by the size of the raise in kilograms, and the proportion predicted
by the fitted model at its posterior mode for each attempt from the weights
declared before its outcome was known. For the women's snatch the mode is
that of the region with the larger posterior mass. Raises of one kilogram are rare, eight to twenty-one
attempts in each combination.}
\label{fig:jump}
\end{figure}

The declared weights are chosen to equalize the chance of success.
Figure~\ref{fig:jump} shows, for second attempts that follow a successful
opener, the proportion that succeeded against the size of the raise. For
raises between two and ten kilograms the proportion lies between $0.58$ and
$0.69$ in all four combinations of sex and lift, with no trend. Third attempts after a successful second attempt decline from $0.42$ to $0.49$ at
two kilograms to $0.26$ to $0.36$ at eight to ten. If raises were chosen without knowledge
of capability, a larger raise would leave a smaller margin, and success
would fall steeply with its size. The flat curve of the second attempt
indicates that athletes raise by more when they have more in reserve.

The outcome of an opener also depends little on how it compares with the
athlete's previous best.
Table~\ref{tab:openprev} classifies openers by their ratio to the
athlete's best made lift in the previous four hundred days. The median
opener is $96$ to $97$ per cent of that best, and its success rate falls
only from $0.88$ to $0.93$ for openers below $88$ per cent of it to $0.71$
to $0.79$ for openers at or above it, in all four combinations. Openers
above the previous best,
declared by athletes who have improved, succeed as often as those just
below it. If the previous best
were the capability, a trial whose scale is a few per cent of the bar would
move from near certainty to near failure across this range. Either the
previous best is a poor measure of the capability of the day, or the
outcome of a lift depends weakly on its margin, and the archive by itself
cannot say which. Section~\ref{sec:results-tau} shows that this is the
central fact about what the data identify.

\begin{table}[t]
\centering
\caption{Openers relative to the athlete's heaviest lift made in the
previous four hundred days, for athlete-days with such a lift: number of
openers in each band of the ratio and proportion that succeeded.}
\label{tab:openprev}
\footnotesize\setlength{\tabcolsep}{3.5pt}
\begin{tabular}{lrrrrrrrr}
\toprule
 & \multicolumn{2}{c}{Men, snatch} & \multicolumn{2}{c}{Men, c\&j} & \multicolumn{2}{c}{Women, snatch} & \multicolumn{2}{c}{Women, c\&j} \\
\cmidrule(lr){2-3}\cmidrule(lr){4-5}\cmidrule(lr){6-7}\cmidrule(lr){8-9}
Opener / previous best & $n$ & made & $n$ & made & $n$ & made & $n$ & made \\
\midrule
below $0.88$ & 191 & 0.88 & 166 & 0.89 & 279 & 0.89 & 230 & 0.93 \\
$0.88$ to $0.91$ & 278 & 0.89 & 234 & 0.86 & 416 & 0.84 & 338 & 0.88 \\
$0.91$ to $0.94$ & 792 & 0.83 & 698 & 0.86 & 987 & 0.80 & 852 & 0.87 \\
$0.94$ to $0.97$ & 2{,}068 & 0.79 & 1{,}733 & 0.81 & 1{,}866 & 0.78 & 1{,}722 & 0.84 \\
$0.97$ to $1.00$ & 2{,}026 & 0.72 & 2{,}078 & 0.76 & 1{,}717 & 0.72 & 1{,}802 & 0.79 \\
$1.00$ to $1.03$ & 450 & 0.72 & 559 & 0.71 & 380 & 0.72 & 463 & 0.78 \\
above $1.03$ & 371 & 0.73 & 381 & 0.77 & 299 & 0.73 & 366 & 0.79 \\
\bottomrule
\end{tabular}
\end{table}

Strength grows with bodyweight more slowly than the classical scaling
argument predicts. A regression of the log best made lift on log
bodyweight and year gives exponents from $0.465$ to $0.509$, well below the two-thirds surface law of
\citet{lietzke1956}.

\subsection{The attempt scale}
\label{sec:results-tau}

\begin{figure}[t]
\centering
\includegraphics[width=0.98\textwidth]{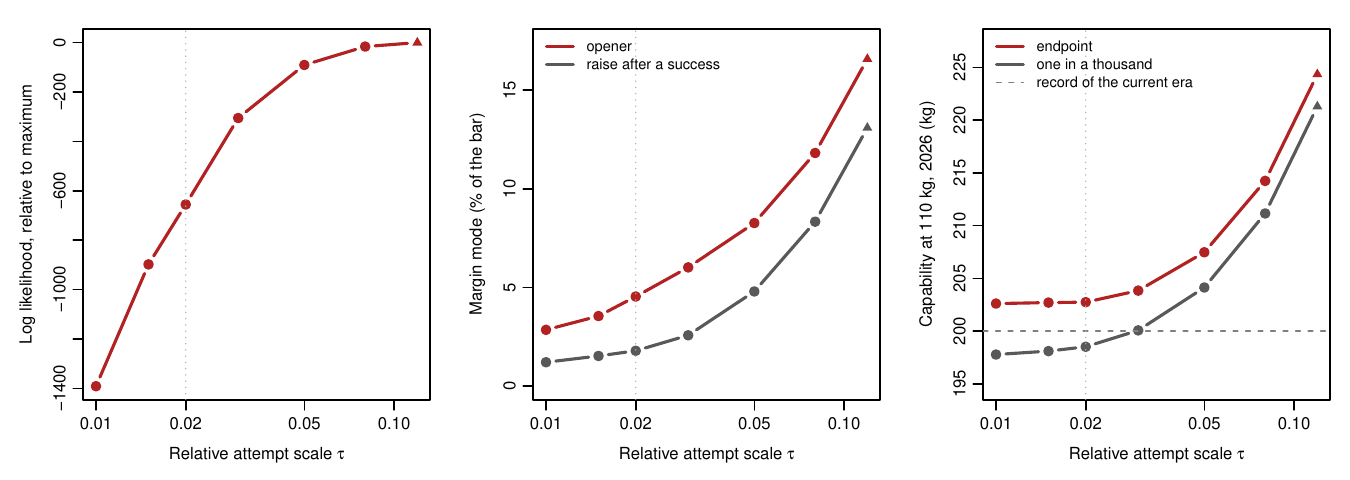}
\caption{The simplified model of Section~\ref{sec:results-tau} for the
men's snatch, along the direction of
Proposition~\ref{prop:ridge}: profile log likelihood relative to its
maximum, modes of the opener margin and of the raise after a success, and
the endpoint and one-in-a-thousand capability at $110$ kilograms of
bodyweight in 2026, against the fixed attempt scale $\tau$ on a log axis.
The triangle is the fit with $\tau$ free. The dotted line marks the
reference anchor $\tau = 0.02$ and the dashed line in the right panel the
local record of the current era.}
\label{fig:ridge}
\end{figure}

Figure~\ref{fig:ridge} follows the direction of
Proposition~\ref{prop:ridge} for the men's snatch, fixing the attempt scale $\tau$ at one to eight per cent of the bar and re-estimating everything else (Section~\ref{sec:s-ridge} of the supplement). It uses a simplified model, without the athlete factor and with one lower margin scale shared across the three situations, cheap enough to trace the whole direction. The profile log likelihood rises without
a maximum in the range that physiology allows, by $1{,}086$ units from
$\tau = 0.01$ to $\tau = 0.03$. The margins grow roughly in step with
$\tau$, as the proposition says they must, while the tail shape stays
between $-0.32$ and $-0.39$. The endpoint at $110$ kilograms of bodyweight
rises from $202.6$ kilograms at $\tau = 0.01$ to $224.3$ in the free fit,
which ends at $\tau = 0.120$ with openers $17$ per cent below
capability. The final model behaves alike. Left free, it ends at
$\tau = 0.144$, $1{,}349$ units of log likelihood above the fit at
$\tau = 0.02$, with openers $17$ per cent below capability, the tail shape
at the upper limit of its support and the men's snatch ceiling infinite.
The archive prefers a description
in which a world record is a lucky lift by an athlete whose capability lies
far above what is declared. That description contradicts the measured variability of maximal strength. The rest of this section fixes $\tau$ at $0.02$, with the other two anchors alongside.

\subsection{Attempts and declarations}
\label{sec:results-decl}

Table~\ref{tab:attempt} gives the posterior for the attempt and declaration
parameters at the reference anchor. With $\tau = 0.02$ an opener two per
cent below capability, three kilograms on a bar of $150$ kilograms,
succeeds about three times in four, and one four per cent below about nine
times in ten. At the same margin between
capability and bar, the second attempt succeeds with $1.4$ to $2.3$ times
the odds of the opener and the third with $1.2$ to $2.9$ times. This reflects partly the nerves of the first lift and partly that an athlete who raises is one whose previous lift went up easily. The low raw
success rate of third attempts therefore comes from where the bar is
placed.

The margin densities differ in shape between the sexes but not in their
averages, and the average opener margin is $2.9$ to $4.2$ per cent of the
bar in all four combinations. The men's opener margin is close to
symmetric about a mode of $3.3$ and $3.6$ per cent, while the women's has modes of $1.8$ and $0.8$ per cent with a long upper tail of conservative openers. After a success men raise to about their capability, while women again show a low mode and a long upper tail. The margin after a miss, the
minority choice, is weakly determined (Section~\ref{sec:s-sens} of the supplement).
Three scales sit at their lower bound of $0.3$ per cent of the bar, among
them the upper scale of the men's raises after a success. At bars of $100$
to $250$ kilograms the bound is $0.3$ to $0.75$ kilograms, below the
resolution of the declared weights, which are whole kilograms. At this anchor the athlete opens
three to four per cent below what he or she can lift, raises to about
capability after a success, and treats a third attempt as a toss-up.

\begin{table}[t]
\centering
\caption{Posterior medians and $95$ per cent intervals for the attempt and
declaration parameters at the reference anchor $\tau = 0.02$: the
attempt-order shifts on the log odds scale, and the modes and scales of the
margin between capability and the declared weight, in per cent of the bar,
for openers and raises after a success.}
\label{tab:attempt}
\footnotesize\setlength{\tabcolsep}{3.5pt}
\begin{tabular}{llrrrr}
\toprule
 & & Men, snatch & Men, c\&j & Women, snatch & Women, c\&j \\
\midrule
$\delta_2$ &  & 0.66 (0.59, 0.72) & 0.37 (0.30, 0.44) & 0.82 (0.75, 0.90) & 0.43 (0.35, 0.50) \\
$\delta_3$ &  & 0.75 (0.68, 0.81) & 0.15 (0.08, 0.22) & 1.06 (0.99, 1.14) & 0.24 (0.16, 0.31) \\
Opener & mode & 3.3 (3.0, 3.5) & 3.6 (3.4, 3.9) & 1.8 (1.5, 2.1) & 0.8 (0.6, 1.0) \\
 & upper scale & 1.7 (1.6, 1.9) & 1.8 (1.6, 1.9) & 4.2 (4.0, 4.4) & 4.6 (4.5, 4.7) \\
 & lower scale & 2.2 (2.1, 2.3) & 2.3 (2.1, 2.4) & 2.0 (1.8, 2.2) & 0.3 (0.3, 0.5) \\
Raise after a success & mode & 1.5 (1.4, 1.7) & 2.0 (1.9, 2.2) & 1.6 (1.2, 2.0) & 0.1 ($-$0.2, 0.4) \\
 & upper scale & 0.3 (0.3, 0.3) & 0.3 (0.3, 0.3) & 3.0 (2.4, 3.9) & 11.4 (8.2, 20.9) \\
 & lower scale & 2.9 (2.9, 3.0) & 3.1 (3.0, 3.1) & 3.2 (3.1, 3.4) & 2.3 (2.2, 2.5) \\
\bottomrule
\end{tabular}
\end{table}

The attempt model assumes that a declared weight says nothing about the
form of the attempt beyond capability and the earlier outcomes. Predicting
each attempt from everything observed before its outcome, the model matches
the success rates by attempt to within $0.035$ in every combination
(Section~\ref{sec:s-sens} of the supplement), but it predicts a decline with the size of
the raise where the observed curve is flat (Figure~\ref{fig:jump}). In the
men's snatch the predicted rate falls from $0.66$ at two kilograms to $0.40$
at eight to ten, against observed rates between $0.60$ and $0.69$. The
margin density of a raise after a success has a lower scale of about three
per cent, so the model reads a large raise partly as a bar above capability and
predicts failure in proportion. A flat predicted curve would
require the outcome of a lift to depend less on its margin, which is
the direction of Figure~\ref{fig:ridge}. The misfit is the visible price of
the anchor.

\subsection{The body and the tail}
\label{sec:results-tail}

Table~\ref{tab:tail} collects the parameters of the capability
distribution. In every combination the body has a long lower tail, which
carries the athlete-days of the weak fields. Its components trade off
against one another and against the location coefficients, so they are less well determined than the distribution they make up. The
threshold lies $1.5$ to $2.0$ scale units above the location, at the $64$th
to $93$rd percentile of the error, and is the least well
determined parameter of the tail. In the men's clean and jerk it lies at the
$64$th percentile, so the tail covers the upper third. The men's tail is bounded with
posterior probability above $0.999$, with shapes
$\xi_s = -0.31$ $(-0.36, -0.25)$ and $\xi_c = -0.40$ $(-0.44, -0.36)$. For
women the shapes are $-0.23$ $(-0.29, 0.05)$ in the snatch and $-0.18$
$(-0.28, 0.14)$ in the clean and jerk, negative with posterior
probabilities $0.96$ and $0.94$. The upper ends of the women's intervals
come from the part of each posterior that sits at shapes near zero, which
for the snatch is the higher-threshold region of
Section~\ref{sec:method-comp}. The athlete correlation $\rho_a$ is $0.90$ to $0.93$, so about ninety per
cent of the variation in capability is persistent within athletes.

\begin{table}[t]
\centering
\caption{Posterior medians and $95$ per cent intervals for the capability
distribution at the reference anchor: the threshold $u_l$ on the scale of
the standardized error, the probability $H_l(u_l)$ that the body places
below it, the scale and shape of the generalized Pareto tail, the
correlation between two competitions of the same athlete, the probability that the tail
has a finite endpoint, and the time trend of the location.}
\label{tab:tail}
\footnotesize\setlength{\tabcolsep}{3.5pt}
\resizebox{\textwidth}{!}{%
\begin{tabular}{lrrrr}
\toprule
 & Men, snatch & Men, c\&j & Women, snatch & Women, c\&j \\
\midrule
$u_l$ & 1.75 (1.12, 2.56) & 2.00 (0.93, 3.26) & 1.47 (0.94, 2.48) & 1.63 (1.01, 3.17) \\
$H_l(u_l)$ & 0.846 (0.805, 0.877) & 0.643 (0.596, 0.723) & 0.855 (0.799, 0.992) & 0.933 (0.857, 0.999) \\
$\sigma_l$ & 0.42 (0.35, 0.53) & 0.71 (0.55, 0.95) & 0.96 (0.19, 1.64) & 0.57 (0.14, 0.85) \\
$\xi_l$ & $-$0.31 ($-$0.36, $-$0.25) & $-$0.40 ($-$0.44, $-$0.36) & $-$0.23 ($-$0.29, 0.05) & $-$0.18 ($-$0.28, 0.14) \\
$\rho_{a,l}$ & 0.906 (0.900, 0.912) & 0.895 (0.889, 0.902) & 0.921 (0.915, 0.926) & 0.929 (0.923, 0.934) \\
$P(\xi_l < 0)$ & 1.00 & 1.00 & 0.96 & 0.94 \\
Location trend (kg per year) & 0.32 (0.26, 0.37) & 0.36 (0.29, 0.44) & 0.60 (0.55, 0.64) & 0.79 (0.74, 0.84) \\
\bottomrule
\end{tabular}}
\end{table}

\subsection{From the record book to capability}
\label{sec:results-ladder}

The difference between capability and the recorded best lift is a posterior
quantity, and at the reference anchor it is within $0.7$ kilograms on
average in all four combinations. Its sign depends on how the day ended.
When a heavier attempt was missed, capability lies $0.5$ to $2.1$ kilograms
above the best made lift, since the miss bounds it from above. When the day
ended on a success, it lies $0.6$ to $2.0$ kilograms below, since the final
attempt was declared at a weight the athlete makes about as often as not.
Of the one per cent of athlete-days with the largest best lift relative to
bodyweight and era, $76$ to $84$ per cent ended on a successful attempt,
against $39$ to $49$ per cent of all athlete-days with a made lift. At this
anchor the record book therefore measures median capability with little
bias on average and overstates it at the top. At a larger anchor it is a
lower bound throughout, since along Figure~\ref{fig:ridge} the gap grows in
step with $\tau$.

Table~\ref{tab:ladder} separates what each layer of the model contributes
to the tail. Every row uses the same distribution of capability, the
normal-mixture body with a generalized Pareto tail above an estimated
threshold, and the same priors, so the rows differ only in what the model is
fitted to. The first row is the record book. It takes the best made lift of each
athlete-day as the capability and the athlete-days as exchangeable, and
since lifts are recorded to the kilogram, each best lift contributes the
probability of its kilogram interval. The second row adds the athlete factor of
Section~\ref{sec:method-athlete} to the same best lifts. The third is the
full model, with the attempts and the declarations. The last row uses no
model for the body. It places a threshold at the $0.9$ quantile of the best
lift from a linear quantile regression on bodyweight and year, fits a
generalized Pareto with a log scale linear in bodyweight to the
exceedances, and takes intervals from a bootstrap over athletes
(Section~\ref{sec:s-direct} of the supplement).

Taken as an exchangeable sample, the record book puts the men's ceiling at
$110$ kilograms of bodyweight at $212$ kilograms in the snatch and $257$
in the clean and jerk, six per cent above the local records of $200$ and
$242$. The check without a body agrees with it to within two
kilograms. Linking the competitions of the same athlete leaves the record
book unable to settle the tail. With the athlete factor the probability of a finite ceiling falls to $0.82$ and $0.79$. The intervals for the shape reach $0.28$. The exchangeable fit treats thirty competitions of one
champion as thirty independent draws from the upper tail. Counted as the
results of one athlete, the champions are a few draws from a population, and a few draws cannot fix where the tail ends. The entries with the
athlete factor for the men's clean and jerk and the women's snatch are
indicative, as their thresholds mix slowly (Section~\ref{sec:s-conv} of the supplement).

The attempts and the declarations fix it. In the full model the men's tail
is bounded with posterior probability above $0.999$, with shapes of $-0.31$ and $-0.40$. The ceilings are $206$ and $245$ kilograms, six and twelve kilograms below the record book, three and one per cent above the local records. The
comparison of capability with the best made lift explains the difference. A
best made lift is capability measured with error, largest on the days that
end with a successful final attempt, which make up most of the top of the
archive. The latent
distribution is free of that error, and its upper tail is shorter than the
tail of the best lifts.

For women the layers disagree in both directions. The record book leaves
the snatch
unbounded with probability $0.16$ when the athlete-days are exchangeable,
and the clean and jerk with probability $0.26$ when the athlete factor is
added. In the full model the ceilings at $86$ kilograms are $143$ and $187$
kilograms, twelve and sixteen per cent above the local records of $127$
and $161$. The check without a body gives $140$ and $187$. In the
snatch the record book places the ceiling seven kilograms above the full
model, at $150$, as it does for men. In the clean and jerk it places the
ceiling below the full
model, at $180$ kilograms, because it estimates a shorter tail, with a
shape of $-0.30$ against $-0.18$, and the full model's tail for that lift is
the least well determined in the paper.

\begin{table}[t]
\centering
\caption{The tail of the capability distribution under successive layers of
the model, with the same mixture body and threshold estimation throughout:
posterior median and $95$ per cent interval of the tail shape, posterior
probability that the tail is bounded, and the posterior median of the endpoint at the reference bodyweight ($110$
kilograms for men, $86$ for women) in 2026 with, in parentheses, the
posterior probability that it lies within ten per cent of the local record
of the current era. The last row uses no body: a quantile-regression
threshold, a generalized Pareto for the exceedances and an athlete
bootstrap, with the share of bootstrap samples with a negative shape and
the point estimate of the endpoint.}
\label{tab:ladder}
\footnotesize\setlength{\tabcolsep}{3.5pt}
\resizebox{\textwidth}{!}{%
\begin{tabular}{llrrrr}
\toprule
 & & Men, snatch & Men, c\&j & Women, snatch & Women, c\&j \\
 & & (record 200) & (record 242) & (record 127) & (record 161) \\
\midrule
Record book & $\xi_l$ & $-$0.23 ($-$0.36, 0.10) & $-$0.21 ($-$0.27, $-$0.14) & $-$0.13 ($-$0.31, 0.15) & $-$0.30 ($-$0.36, $-$0.21) \\
 & $P(\xi_l < 0)$ & 0.93 & 1.00 & 0.84 & 1.00 \\
 & ceiling & 212 (0.84) & 257 (0.94) & 150 (0.02) & 180 (0.07) \\
Record book, athlete factor & $\xi_l$ & $-$0.24 ($-$0.57, 0.28) & $-$0.11 ($-$0.49, 0.28) & $-$0.18 ($-$0.22, $-$0.14) & $-$0.15 ($-$0.56, 0.32) \\
 & $P(\xi_l < 0)$ & 0.82 & 0.79 & 1.00 & 0.74 \\
 & ceiling & 219 (0.53) & 307 (0.08) & 151 (0.00) & 204 (0.00) \\
Full model & $\xi_l$ & $-$0.31 ($-$0.36, $-$0.25) & $-$0.40 ($-$0.44, $-$0.36) & $-$0.23 ($-$0.29, 0.05) & $-$0.18 ($-$0.28, 0.14) \\
 & $P(\xi_l < 0)$ & 1.00 & 1.00 & 0.96 & 0.94 \\
 & ceiling & 206 (1.00) & 245 (1.00) & 143 (0.11) & 187 (0.02) \\
No body (check) & $\xi_l$ & $-$0.24 ($-$0.34, $-$0.16) & $-$0.22 ($-$0.31, $-$0.15) & $-$0.32 ($-$0.43, $-$0.20) & $-$0.21 ($-$0.33, $-$0.08) \\
 & $P(\xi_l < 0)$ & 1.00 & 1.00 & 1.00 & 1.00 \\
 & ceiling & 212 & 255 & 140 & 187 \\
\bottomrule
\end{tabular}}
\end{table}

\subsection{Dependence between the lifts}
\label{sec:results-dep}

\begin{figure}[t]
\centering
\includegraphics[width=0.9\textwidth]{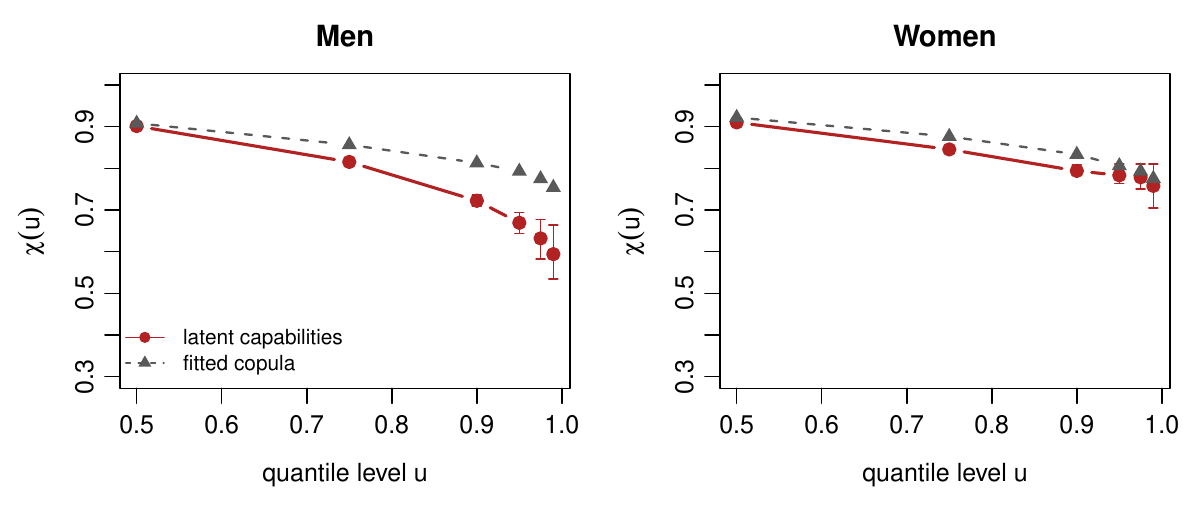}
\caption{Extremal dependence between the lifts at observable levels:
$\chi(u)$ computed on the posterior draws of the latent capabilities, with
intervals over the draws, against the value implied by the fitted copula.}
\label{fig:chimodel}
\end{figure}

The copula gives a correlation $\rho$ of $0.96$
for men and $0.97$ for women. Its tail parameter $\psi$ is $0.15$ for men, a
$t$ copula with seven to ten degrees of freedom, and the Gaussian case
$\psi = 0$, under which the two lifts would be asymptotically independent,
has posterior probability below $10^{-11}$. For women $\psi$ is $0.10$ $(0.05, 0.10)$,
ten to twenty degrees of freedom, and the Gaussian case has posterior
probability $10^{-5}$. On the posterior draws of the latent
capabilities, the probability $\chi(u)$ that the clean and jerk
exceeds its $u$-quantile given that the snatch exceeds its own is $0.59$
$(0.54, 0.66)$ at $u = 0.99$ for men and $0.76$ $(0.71, 0.81)$ for women. It declines slowly and stays well above zero (Figure~\ref{fig:chimodel}). The fitted copula gives $0.75$ and $0.78$ at
that level. For women the two agree. For men the latent $\chi(u)$ is still
falling at $u = 0.99$, and a $t$ copula, whose $\chi(u)$ settles at a
positive limit, may overstate the dependence at the top. The
latent draws also carry independent posterior uncertainty about each
capability, which dilutes the latent $\chi(u)$, so for men
the dependence at the top lies between the two curves. The
competition total therefore loses almost nothing to the need to be extreme
in both lifts on the same day. At one in ten thousand athlete-days the
men's total at $110$ kilograms is $445.8$
$(442.7, 449.0)$, $0.3$ kilograms below the sum of the two marginal levels on the same draws, and the women's deficit at $86$ kilograms is $0.4$.

\subsection{Predictions}
\label{sec:results-pred}

Table~\ref{tab:quant} gives the capability quantiles in 2026 at the
reference anchor, with the local record of the current era, the heaviest
lift made since 2019 within five kilograms of each bodyweight. Records of the current era are used
because the location drifts with time. Lifts of 2017 and 2018 exceed them
at seven of the forty combinations of bodyweight and lift, by one to four
kilograms. The largest, a clean and jerk of $233$ kilograms at $94.0$
kilograms of bodyweight in 2017, lies $5.2$ kilograms above the ceiling of
2017 at that bodyweight, about two per cent, within the variation of a
single attempt, whose momentary maximum has a standard deviation of $3.6$
per cent of capability. Up to $110$ kilograms of bodyweight a
local record is the maximum of $650$ to $1{,}150$ athlete-days, so under
the model it should sit near or somewhat below the one-in-a-thousand level. At $110$
kilograms that level is $199.3$ kilograms in the snatch and $241.0$ in the
clean and jerk, against records of $200$ and $242$. The record
lies within five kilograms of that level at twelve of the sixteen
combinations of bodyweight and lift. For women the record does so at
sixteen of the twenty combinations and lies furthest below the level at
$86$ kilograms, by five and seven kilograms. These levels describe the strongest plausible competitor of each size
today. Competitions simulated from the posterior, attempts included,
reproduce the local records at $36$ of the $40$ combinations (Section~\ref{sec:s-sens} of
the supplement). The exceptions are the men's snatch at $61$ kilograms, one
kilogram below the simulated range, and three of the four women's records at the two heaviest bodyweights, which lie two to five kilograms above it.

\begin{table}[t]
\centering
\caption{Posterior medians and $95$ per cent intervals for capability
quantiles in 2026 at the reference anchor, in kilograms, by sex and
bodyweight, the posterior median of the ceiling, and the local record of
the current era (the heaviest lift made since 2019 within five kilograms of
the bodyweight). The $0.999$ level is exceeded by one athlete-day in a
thousand at that bodyweight.}
\label{tab:quant}
\footnotesize\setlength{\tabcolsep}{3.5pt}
\resizebox{\textwidth}{!}{%
\begin{tabular}{rrrrrrrrr}
\toprule
Bodyweight & \multicolumn{4}{c}{Snatch} & \multicolumn{4}{c}{Clean and jerk} \\
\cmidrule(lr){2-5}\cmidrule(lr){6-9}
 & $q_{0.99}$ & $q_{0.999}$ & Ceiling & Record & $q_{0.99}$ & $q_{0.999}$ & Ceiling & Record \\
\midrule
\multicolumn{9}{l}{\textit{Men}} \\
61 & 141.3 (140.2, 142.4) & 146.6 (145.2, 148.0) & 152 & 146 & 173.2 (172.1, 174.5) & 178.1 (176.8, 179.6) & 181 & 183 \\
67 & 151.1 (150.1, 152.1) & 156.7 (155.4, 158.1) & 162 & 160 & 185.0 (184.0, 186.1) & 190.2 (188.9, 191.6) & 193 & 197 \\
73 & 159.5 (158.6, 160.5) & 165.4 (164.1, 166.8) & 171 & 169 & 195.2 (194.2, 196.2) & 200.5 (199.3, 201.9) & 204 & 209 \\
81 & 169.1 (168.1, 170.0) & 175.2 (173.9, 176.7) & 181 & 182 & 206.7 (205.6, 207.7) & 212.2 (211.0, 213.7) & 216 & 211 \\
89 & 177.0 (176.0, 178.0) & 183.4 (182.1, 185.0) & 190 & 188 & 216.2 (215.1, 217.4) & 221.9 (220.6, 223.6) & 226 & 229 \\
96 & 182.8 (181.9, 183.9) & 189.5 (188.1, 191.1) & 196 & 188 & 223.3 (222.2, 224.5) & 229.2 (227.8, 230.9) & 233 & 233 \\
102 & 187.2 (186.2, 188.3) & 194.0 (192.6, 195.7) & 201 & 190 & 228.6 (227.5, 229.9) & 234.6 (233.1, 236.4) & 239 & 233 \\
110 & 192.3 (191.2, 193.4) & 199.3 (197.8, 201.0) & 206 & 200 & 234.8 (233.5, 236.1) & 241.0 (239.3, 242.7) & 245 & 242 \\
130 & 201.7 (200.4, 203.2) & 209.2 (207.5, 211.1) & 216 & 204 & 246.5 (244.9, 248.3) & 253.0 (251.0, 255.2) & 257 & 250 \\
150 & 208.0 (206.1, 210.1) & 215.7 (213.6, 218.3) & 223 & 217 & 254.5 (252.2, 257.0) & 261.2 (258.6, 264.2) & 266 & 253 \\
\addlinespace
\multicolumn{9}{l}{\textit{Women}} \\
49 & 95.4 (94.4, 96.5) & 100.8 (99.4, 102.4) & 109 & 99 & 120.7 (119.4, 122.2) & 128.4 (126.6, 130.7) & 144 & 126 \\
55 & 102.7 (101.8, 103.7) & 108.4 (107.1, 109.9) & 117 & 111 & 129.5 (128.3, 130.8) & 137.6 (135.9, 139.8) & 154 & 141 \\
59 & 106.8 (105.9, 107.8) & 112.7 (111.4, 114.2) & 122 & 117 & 134.6 (133.4, 135.9) & 143.0 (141.2, 145.2) & 160 & 149 \\
64 & 111.3 (110.4, 112.3) & 117.5 (116.1, 119.0) & 127 & 120 & 140.2 (139.0, 141.5) & 148.9 (147.1, 151.2) & 166 & 151 \\
71 & 116.6 (115.7, 117.7) & 123.1 (121.7, 124.7) & 133 & 125 & 146.9 (145.6, 148.3) & 156.0 (154.2, 158.4) & 174 & 156 \\
76 & 119.8 (118.8, 121.0) & 126.5 (125.1, 128.1) & 137 & 125 & 151.1 (149.7, 152.5) & 160.4 (158.5, 162.9) & 179 & 161 \\
81 & 122.7 (121.6, 123.8) & 129.5 (128.0, 131.2) & 140 & 127 & 154.8 (153.3, 156.3) & 164.3 (162.4, 166.8) & 183 & 161 \\
86 & 125.2 (124.0, 126.4) & 132.1 (130.6, 133.8) & 143 & 127 & 158.1 (156.6, 159.7) & 167.8 (165.8, 170.3) & 187 & 161 \\
100 & 130.7 (129.4, 132.2) & 138.0 (136.4, 139.7) & 149 & 145 & 165.6 (163.8, 167.6) & 175.7 (173.6, 178.5) & 196 & 178 \\
120 & 135.8 (134.0, 138.0) & 143.5 (141.4, 145.9) & 155 & 140 & 173.3 (170.8, 176.0) & 183.9 (180.9, 187.3) & 205 & 185 \\
\bottomrule
\end{tabular}}
\end{table}

Table~\ref{tab:ceiling} reports the ceiling. For men the probability of a
finite ceiling is above $0.999$ in both disciplines and at every bodyweight.
The probability that it lies within ten per cent of the local record is
at least $0.998$. At
$110$ kilograms of bodyweight the median ceiling is $206$ kilograms in the
snatch and $245$ in the clean and jerk, three and one per cent above the
records of $200$ and $242$. Across the bodyweight range the men's
ceilings lie at most six per cent above the records. At five of the twenty
men's combinations the ceiling lies up to $2.4$ per cent below the record,
and in all five the record was lifted within $1.4$ kilograms of the upper
edge of the five-kilogram window. Against the ceiling at the lifter's own
bodyweight every local record lies below it, the closest being the men's
clean and jerk record of $209$ kilograms, lifted at $76.7$ kilograms of
bodyweight and $0.6$ kilograms below the ceiling there. For women the ceilings
lie $3$ to $16$ per cent above the local records across the bodyweights of
Table~\ref{tab:quant}, furthest at $86$ kilograms. \citet{einmahl2026} found the same pattern for the women's
$100$ metres and attributed it in part to the smaller number of women in
their data.

Fitted to the whole archive since 1998, the model places the ceilings
higher, at $210$ and $250$ kilograms for men and $157$ and $196$ for women
(Section~\ref{sec:s-archive} of the supplement). The heaviest women's lifts at $86$
kilograms, $138$ in the snatch and $166$ in the clean and jerk, date from
2009 and 2012 (Section~\ref{sec:data}). The trend
carries the tail they imply to 2026.

\begin{table}[t]
\centering
\caption{The ceiling. Posterior probability that the capability
distribution has a finite endpoint, that the endpoint lies within ten,
twenty-five and fifty per cent of the heaviest lift made since 2019 by a
competitor within five kilograms of the bodyweight (the local record, in
kilograms), and the posterior median of the endpoint in 2026 at the
reference anchor $\tau = 0.02$. The last three columns give the posterior
mode of the endpoint at the three anchors for the reference bodyweight of
each sex, for the women's snatch at the mode of the region with the larger
posterior mass.}
\label{tab:ceiling}
\footnotesize\setlength{\tabcolsep}{3.5pt}
\begin{tabular}{llrrrrrrrrrr}
\toprule
 & & & & & & & & & \multicolumn{3}{c}{Mode at $\tau$} \\
\cmidrule(lr){10-12}
Sex & Lift & Bodyweight & Record & $P(\text{finite})$ & $+10\%$ & $+25\%$ & $+50\%$ &
Ceiling & $0.01$ & $0.02$ & $0.03$ \\
\midrule
Men & Snatch & 73 & 169 & 1.00 & 1.00 & 1.00 & 1.00 & 171 &  &  &  \\
Men & Snatch & 89 & 188 & 1.00 & 1.00 & 1.00 & 1.00 & 190 &  &  &  \\
Men & Snatch & 110 & 200 & 1.00 & 1.00 & 1.00 & 1.00 & 206 & 207 & 206 & 211 \\
Men & Snatch & 150 & 217 & 1.00 & 1.00 & 1.00 & 1.00 & 223 &  &  &  \\
Men & C\&J & 73 & 209 & 1.00 & 1.00 & 1.00 & 1.00 & 204 &  &  &  \\
Men & C\&J & 89 & 229 & 1.00 & 1.00 & 1.00 & 1.00 & 226 &  &  &  \\
Men & C\&J & 110 & 242 & 1.00 & 1.00 & 1.00 & 1.00 & 245 & 247 & 245 & 248 \\
Men & C\&J & 150 & 253 & 1.00 & 1.00 & 1.00 & 1.00 & 266 &  &  &  \\
Women & Snatch & 59 & 117 & 0.96 & 0.85 & 0.92 & 0.93 & 122 &  &  &  \\
Women & Snatch & 71 & 125 & 0.96 & 0.80 & 0.91 & 0.93 & 133 &  &  &  \\
Women & Snatch & 86 & 127 & 0.96 & 0.11 & 0.89 & 0.93 & 143 & 143 & 142 & 146 \\
Women & Snatch & 120 & 140 & 0.96 & 0.35 & 0.90 & 0.93 & 155 &  &  &  \\
Women & C\&J & 59 & 149 & 0.94 & 0.67 & 0.90 & 0.92 & 160 &  &  &  \\
Women & C\&J & 71 & 156 & 0.94 & 0.36 & 0.87 & 0.92 & 174 &  &  &  \\
Women & C\&J & 86 & 161 & 0.94 & 0.02 & 0.83 & 0.92 & 187 & 191 & 187 & 187 \\
Women & C\&J & 120 & 185 & 0.94 & 0.44 & 0.88 & 0.92 & 205 &  &  &  \\
\bottomrule
\end{tabular}
\end{table}

Within the anchors that physiology allows, the ceiling barely moves. Between $\tau = 0.01$ and $0.03$ the posterior modes of the
ceilings at the reference bodyweight move by at most five kilograms
(Table~\ref{tab:ceiling}) and the one-in-a-thousand level of the men's
snatch by $4.5$. The attempt-order shifts absorb most of the change in $\tau$, which alters how a record is read. The mean difference between capability and best made lift in the men's snatch is $-1.4$
kilograms at $\tau = 0.01$, $-0.4$ at $0.02$ and $0.7$ at $0.03$. A simulation study (Section~\ref{sec:s-ridge} of the supplement) shows the same on data
generated at $\tau = 0.02$. Refitted at the true anchor, the model recovers
the tail. Refitted at $0.01$ or $0.03$, it moves the attempt-order shifts and margins, while the one-in-a-thousand levels and ceilings move by at most $1.8$ kilograms on average.

\section{Discussion}
\label{sec:discussion}

The men's results sit close to those obtained for other sports from record
books. \citet{einmahl2008} found the men's athletics records within a few
per cent of their estimated limits, and \citet{einmahl2026} found the men's
$100$ metres record within hundredths of a second of its limit with a much
larger margin for women. Here the men's ceiling lies at most six per cent
above the heaviest lifts of the current era and the women's $3$ to $16$
per cent above theirs, across the bodyweights of Table~\ref{tab:quant}. The same distribution fitted to the best lifts, as
those analyses do, places the men's ceilings six and twelve kilograms
higher than the full model does, a body-free generalized Pareto fit six and ten kilograms higher. Table~\ref{tab:ladder} shows where the difference comes from. The record book reads the luck of a made lift as capability, which lengthens the tail. Once the competitions of the same athlete are linked it no longer settles where the tail ends.
The extremal dependence between the two lifts resembles that found by
\citet{adam2012} in bivariate Olympic data.

The treatment of the design places the paper beside the work on adaptive
testing and informative censoring \citep{mislevy2000,little2019}. When a
latent quantity is probed by challenges the subject chooses, a model that
conditions on the challenges loses the identification of the measurement scale. A model for the choice restores most of it.
Proposition~\ref{prop:ridge} names what remains open, the position of the
latent scale relative to the challenges, which the paper settles with
outside knowledge of how much a lift varies from one attempt to the next. The extreme value mixture model of \citet{behrens2004} and
\citet{nascimento2012} carries over to a latent variable model with a gain.
With the latent quantities integrated out, the marginal likelihood is
smooth in the threshold, which removes the difficulty of threshold
estimation with observed data. Estimating the threshold also matters. Held at a fixed quantile, the threshold moves the record-book
ceilings by up to ten kilograms (Section~\ref{sec:s-splice} of the supplement).

Several limitations qualify the claims. The first is the anchor. At
$\tau = 0.02$ the momentary maximum has a standard deviation of $3.6$ per
cent of capability, close to the between-session coefficient of variation
of $3.3$ per cent reported by \citet{grgic2020}. The within-day variation that $\tau$ represents is plausibly smaller. Between
$\tau = 0.01$ and $\tau = 0.03$ the posterior modes of the ceilings move
by at most five kilograms (Table~\ref{tab:ceiling}). The anchor acts mainly on the attempt-order shifts and on how a successful record is read. The attempt model is logistic in the margin with a single scale. It predicts a decline of success with the size of a raise that the data do not show. A floor on the
miss probability would remove the misfit, but at a fixed $\tau$ it drifts
along the direction of Figure~\ref{fig:ridge}, where the data cannot tell a
floor from a shift of the capability scale. A shared error in the athlete's assessment of the day's capability improves the fit. It moves the one-in-a-thousand levels by at most $1.3$ kilograms and the ceilings by at most $2.4$, except at the higher-threshold mode of the women's snatch, whose endpoint responds to any change of a shape that sits near zero (Section~\ref{sec:s-sens} of the supplement). The model also understates the very top of the best
lifts. In competitions simulated from the posterior the $0.99$ quantile of
the best made lifts, relative to bodyweight and era, is lower than observed
in at least $95$ per cent of the simulations for every sex and lift. Among the top one per cent of best lifts $54$ to $64$ per cent end on a
made attempt, against $76$ to $84$ per cent in the data (Section~\ref{sec:s-sens} of the
supplement). The model reads some of these made final attempts as luck,
which pulls the inferred capability of the strongest days down. The ceilings are therefore more likely conservative than generous, and the difference from the record book in Table~\ref{tab:ladder} is if anything an upper bound on its bias.

The model also leaves out some structure. The choice of weight responds to
the competitive situation, whether a medal is at stake and what rivals have
declared, which the model does not see. Nor does it describe the choice
between repeating and raising a weight after a miss. The athlete factor is constant over a career,
where a trajectory per athlete, as in \citet{griffin2022}, would follow improvement
and decline. Fitted to the whole archive rather than the years since
2017, the model raises the men's ceilings by four and five kilograms, the women's by fourteen and nine (Section~\ref{sec:s-archive} of
the supplement). Bodyweight is measured at the weigh-in, so the model
gives the capability of an athlete who makes a given weight, which suits
questions about competition better than about physiology.

Weightlifting records are declared before they are lifted. Read through a measurement model, the declarations locate the limit that the records only bound. The capability distribution could also serve the
sport, by scoring a lift through its position in the distribution at the
athlete's own bodyweight. Such a score would not depend on the records it
is compared with. It would have the properties that \citet{spearing2021} established for swimming.

\section*{Funding}
Supported by Spanish State Research Agency grant \mbox{PID2023-153222OB-I00}.

\section*{Code and data}
The R and C++ code and the processed attempt-level data that reproduce
every table and figure of the paper and of the supplement are available at
\url{https://github.com/manueleleonelli/weightlifting-capability}.

\section*{Statement on AI use}
The author used Claude (Anthropic), a large language model, as an assistant
during the research. It was used to write and debug the R and C++ code of
the analysis under the author's direction, to run and check computations,
to check the numbers quoted in the text against the outputs of the code, and
to edit the text for clarity and length. The author checked all code and
results and takes full responsibility for the paper.

\bibliographystyle{plainnat-nc}
\bibliography{weightlifting}

\begin{thebibliography}{45}
\providecommand{\natexlab}[1]{#1}
\providecommand{\url}[1]{\texttt{#1}}
\expandafter\ifx\csname urlstyle\endcsname\relax
  \providecommand{\doi}[1]{doi: #1}\else
  \providecommand{\doi}{doi: \begingroup \urlstyle{rm}\Url}\fi

\bibitem[Aarssen and de~Haan(1994)]{aarssen1994}
Karin Aarssen and Laurens de~Haan.
\newblock On the maximal life span of humans.
\newblock \emph{Mathematical Population Studies}, 4:\penalty0 259--281, 1994.

\bibitem[Adam and Tawn(2012)]{adam2012}
M.~B. Adam and J.~A. Tawn.
\newblock Bivariate extreme analysis of {O}lympic swimming data.
\newblock \emph{Journal of Statistical Theory and Practice}, 6:\penalty0
  510--523, 2012.

\bibitem[Andersen(1970)]{andersen1970}
Erling~B. Andersen.
\newblock Asymptotic properties of conditional maximum-likelihood estimators.
\newblock \emph{Journal of the Royal Statistical Society, Series B},
  32:\penalty0 283--301, 1970.

\bibitem[Behrens et~al.(2004)Behrens, Lopes, and Gamerman]{behrens2004}
Cibele~N. Behrens, Hedibert~F. Lopes, and Dani Gamerman.
\newblock Bayesian analysis of extreme events with threshold estimation.
\newblock \emph{Statistical Modelling}, 4:\penalty0 227--244, 2004.

\bibitem[Beirlant et~al.(2007)Beirlant, Guillou, Dierckx, and
  Fils-Villetard]{beirlant2007}
Jan Beirlant, Armelle Guillou, Goedele Dierckx, and Am{\'e}lie Fils-Villetard.
\newblock Estimation of the extreme value index and extreme quantiles under
  random censoring.
\newblock \emph{Extremes}, 10:\penalty0 151--174, 2007.

\bibitem[Berthelot et~al.(2008)Berthelot, Thibault, Tafflet, Escolano,
  El~Helou, Jouven, Hermine, and Toussaint]{berthelot2008}
Geoffroy Berthelot, Val{\'e}rie Thibault, Muriel Tafflet, Sylvie Escolano, Nour
  El~Helou, Xavier Jouven, Olivier Hermine, and Jean-Fran{\c c}ois Toussaint.
\newblock The citius end: world records progression announces the completion of
  a brief ultra-physiological quest.
\newblock \emph{PLoS ONE}, 3:\penalty0 e1552, 2008.

\bibitem[Coles(2001)]{coles2001}
Stuart Coles.
\newblock \emph{An Introduction to Statistical Modeling of Extreme Values}.
\newblock Springer, London, 2001.

\bibitem[Coles et~al.(1999)Coles, Heffernan, and Tawn]{coles1999}
Stuart Coles, Janet Heffernan, and Jonathan Tawn.
\newblock Dependence measures for extreme value analyses.
\newblock \emph{Extremes}, 2:\penalty0 339--365, 1999.

\bibitem[Croucher(1984)]{croucher1984}
John~S. Croucher.
\newblock An analysis of world weightlifting records.
\newblock \emph{Research Quarterly for Exercise and Sport}, 55:\penalty0
  285--288, 1984.

\bibitem[Davison and Smith(1990)]{davison1990}
A.~C. Davison and R.~L. Smith.
\newblock Models for exceedances over high thresholds.
\newblock \emph{Journal of the Royal Statistical Society, Series B},
  52:\penalty0 393--442, 1990.

\bibitem[do~Nascimento et~al.(2012)do~Nascimento, Gamerman, and
  Lopes]{nascimento2012}
Fernando~F. do~Nascimento, Dani Gamerman, and Hedibert~F. Lopes.
\newblock A semiparametric {B}ayesian approach to extreme value estimation.
\newblock \emph{Statistics and Computing}, 22:\penalty0 661--675, 2012.

\bibitem[Einmahl and He(2026)]{einmahl2026}
John H.~J. Einmahl and Yi~He.
\newblock Accurate estimates of ultimate 100-meter records.
\newblock \emph{Extremes}, 29:\penalty0 671--685, 2026.

\bibitem[Einmahl and Magnus(2008)]{einmahl2008}
John H.~J. Einmahl and Jan~R. Magnus.
\newblock Records in athletics through extreme-value theory.
\newblock \emph{Journal of the American Statistical Association}, 103:\penalty0
  1382--1391, 2008.

\bibitem[Einmahl and Smeets(2011)]{einmahl2011}
John H.~J. Einmahl and Sander G. W.~R. Smeets.
\newblock Ultimate 100-m world records through extreme-value theory.
\newblock \emph{Statistica Neerlandica}, 65:\penalty0 32--42, 2011.

\bibitem[Einmahl et~al.(2008)Einmahl, Fils-Villetard, and
  Guillou]{einmahl2008censoring}
John H.~J. Einmahl, Am{\'e}lie Fils-Villetard, and Armelle Guillou.
\newblock Statistics of extremes under random censoring.
\newblock \emph{Bernoulli}, 14:\penalty0 207--227, 2008.

\bibitem[Ford et~al.(2000)Ford, Detterline, Ho, and Cao]{ford2000}
Lincoln~E. Ford, A.~J. Detterline, K.~K. Ho, and W.~Cao.
\newblock Gender- and height-related limits of muscle strength in world
  weightlifting champions.
\newblock \emph{Journal of Applied Physiology}, 89:\penalty0 1061--1064, 2000.

\bibitem[Fraga~Alves et~al.(2017)Fraga~Alves, Neves, and
  Ros{\'a}rio]{alves2017}
Isabel Fraga~Alves, Cl{\'a}udia Neves, and Pedro Ros{\'a}rio.
\newblock A general estimator for the right endpoint with an application to
  supercentenarian women's records.
\newblock \emph{Extremes}, 20:\penalty0 199--237, 2017.

\bibitem[Grgic et~al.(2020)Grgic, Lazinica, Schoenfeld, and Pedisic]{grgic2020}
Jozo Grgic, Bruno Lazinica, Brad~J. Schoenfeld, and Zeljko Pedisic.
\newblock Test-retest reliability of the one-repetition maximum ({1RM})
  strength assessment: a systematic review.
\newblock \emph{Sports Medicine -- Open}, 6:\penalty0 31, 2020.

\bibitem[Griffin et~al.(2022)Griffin, Hinoveanu, and Hopker]{griffin2022}
Jim~E. Griffin, Laurentiu~C. Hinoveanu, and James~G. Hopker.
\newblock Bayesian modelling of elite sporting performance with large
  databases.
\newblock \emph{Journal of Quantitative Analysis in Sports}, 18:\penalty0
  253--267, 2022.

\bibitem[Griffin et~al.(2026)Griffin, Spyropoulou, and Hopker]{griffin2026}
Jim~E. Griffin, Maria-Zafeiria Spyropoulou, and James~G. Hopker.
\newblock Modelling between- and within-season trajectories in elite athletic
  performance data.
\newblock \emph{Journal of the Royal Statistical Society, Series C}, 2026.
\newblock Article qlag002.

\bibitem[Gustafson(2015)]{gustafson2015}
Paul Gustafson.
\newblock \emph{Bayesian Inference for Partially Identified Models: Exploring
  the Limits of Limited Data}.
\newblock Chapman and Hall/CRC, Boca Raton, 2015.

\bibitem[Heffernan and Tawn(2004)]{heffernan2004}
Janet~E. Heffernan and Jonathan~A. Tawn.
\newblock A conditional approach for multivariate extreme values.
\newblock \emph{Journal of the Royal Statistical Society, Series B},
  66:\penalty0 497--546, 2004.

\bibitem[Jensen et~al.(2009)Jensen, Shirley, and Wyner]{jensen2009}
Shane~T. Jensen, Kenneth~E. Shirley, and Abraham~J. Wyner.
\newblock Bayesball: a {B}ayesian hierarchical model for evaluating fielding in
  major league baseball.
\newblock \emph{The Annals of Applied Statistics}, 3:\penalty0 491--520, 2009.

\bibitem[Kolliari-Turner et~al.(2021)Kolliari-Turner, Oliver, Lima, Mills,
  Wang, Pitsiladis, and Guppy]{kolliari2021}
Alexander Kolliari-Turner, Brian Oliver, Giscard Lima, John~P. Mills, Guan
  Wang, Yannis Pitsiladis, and Fergus~M. Guppy.
\newblock Doping practices in international weightlifting: analysis of
  sanctioned athletes/support personnel from 2008 to 2019 and retesting of
  samples from the 2008 and 2012 {O}lympic {G}ames.
\newblock \emph{Sports Medicine -- Open}, 7:\penalty0 4, 2021.

\bibitem[Ledford and Tawn(1996)]{ledford1996}
Anthony~W. Ledford and Jonathan~A. Tawn.
\newblock Statistics for near independence in multivariate extreme values.
\newblock \emph{Biometrika}, 83:\penalty0 169--187, 1996.

\bibitem[Leonelli(2026)]{leonelli2026exceptional}
Manuele Leonelli.
\newblock How exceptional was the {B}ig {T}hree era? {E}xtremes and persistence
  in men's professional tennis.
\newblock arXiv:2608.27362, 2026.

\bibitem[Lietzke(1956)]{lietzke1956}
M.~H. Lietzke.
\newblock Relation between weightlifting totals and body weight.
\newblock \emph{Science}, 124:\penalty0 486--487, 1956.

\bibitem[Little and Rubin(2019)]{little2019}
Roderick J.~A. Little and Donald~B. Rubin.
\newblock \emph{Statistical Analysis with Missing Data}.
\newblock Wiley, Hoboken, NJ, third edition, 2019.

\bibitem[Manski(2003)]{manski2003}
Charles~F. Manski.
\newblock \emph{Partial Identification of Probability Distributions}.
\newblock Springer, New York, 2003.

\bibitem[McNeil and Frey(2000)]{mcneil2000}
Alexander~J. McNeil and R{\"u}diger Frey.
\newblock Estimation of tail-related risk measures for heteroscedastic
  financial time series: an extreme value approach.
\newblock \emph{Journal of Empirical Finance}, 7:\penalty0 271--300, 2000.

\bibitem[Mislevy and Chang(2000)]{mislevy2000}
Robert~J. Mislevy and Hua-Hua Chang.
\newblock Does adaptive testing violate local independence?
\newblock \emph{Psychometrika}, 65:\penalty0 149--156, 2000.

\bibitem[Nguyen et~al.(2026)Nguyen, Smoliga, Lindaman, and
  Deshpande]{nguyen2026}
Paul-Hieu~V. Nguyen, James~M. Smoliga, Benton Lindaman, and Sameer~K.
  Deshpande.
\newblock Quantifying the limits of human athletic performance: a {B}ayesian
  analysis of elite decathletes.
\newblock \emph{Journal of Quantitative Analysis in Sports}, 2026.
\newblock \doi{10.1515/jqas-2026-0026}.

\bibitem[Rasch(1960)]{rasch1960}
Georg Rasch.
\newblock \emph{Probabilistic Models for Some Intelligence and Attainment
  Tests}.
\newblock Danmarks Paedagogiske Institut, Copenhagen, 1960.

\bibitem[Robinson and Tawn(1995)]{robinson1995}
M.~E. Robinson and J.~A. Tawn.
\newblock Statistics for exceptional athletics records.
\newblock \emph{Journal of the Royal Statistical Society, Series C},
  44:\penalty0 499--511, 1995.

\bibitem[Rubin(1976)]{rubin1976}
Donald~B. Rubin.
\newblock Inference and missing data.
\newblock \emph{Biometrika}, 63:\penalty0 581--592, 1976.

\bibitem[Scarrott and MacDonald(2012)]{scarrott2012}
Carl Scarrott and Anna MacDonald.
\newblock A review of extreme value threshold estimation and uncertainty
  quantification.
\newblock \emph{REVSTAT Statistical Journal}, 10:\penalty0 33--60, 2012.

\bibitem[Sinclair(1985)]{sinclair1985}
Roy~G. Sinclair.
\newblock Normalizing the performances of athletes in {O}lympic weightlifting.
\newblock \emph{Canadian Journal of Applied Sport Sciences}, 10:\penalty0
  94--98, 1985.

\bibitem[Spearing et~al.(2021)Spearing, Tawn, Irons, Paulden, and
  Bennett]{spearing2021}
Harry Spearing, Jonathan Tawn, David Irons, Tim Paulden, and Grace Bennett.
\newblock Ranking, and other properties, of elite swimmers using extreme value
  theory.
\newblock \emph{Journal of the Royal Statistical Society, Series A},
  184:\penalty0 368--395, 2021.

\bibitem[Spearing et~al.(2023)Spearing, Tawn, Irons, and Paulden]{spearing2023}
Jess Spearing, Jonathan~A. Tawn, David Irons, and Tim Paulden.
\newblock A framework for statistical modelling of the extremes of longitudinal
  data, applied to elite swimming.
\newblock arXiv:2306.12419, 2023.

\bibitem[Stephenson and Tawn(2013)]{stephenson2013}
Alec~G. Stephenson and Jonathan~A. Tawn.
\newblock Determining the best track performances of all time using a
  conceptual population model for athletics records.
\newblock \emph{Journal of Quantitative Analysis in Sports}, 9:\penalty0
  67--76, 2013.

\bibitem[ter Braak and Vrugt(2008)]{terbraak2008}
Cajo J.~F. ter Braak and Jasper~A. Vrugt.
\newblock Differential evolution {M}arkov chain with snooker updater and fewer
  chains.
\newblock \emph{Statistics and Computing}, 18\penalty0 (4):\penalty0 435--446,
  2008.

\bibitem[Travis et~al.(2021)Travis, Zourdos, and Bazyler]{travis2021}
S.~Kyle Travis, Michael~C. Zourdos, and Caleb~D. Bazyler.
\newblock Weight selection attempts of elite classic powerlifters.
\newblock \emph{Perceptual and Motor Skills}, 128:\penalty0 507--521, 2021.

\bibitem[Vehtari et~al.(2021)Vehtari, Gelman, Simpson, Carpenter, and
  B{\"u}rkner]{vehtari2021}
Aki Vehtari, Andrew Gelman, Daniel Simpson, Bob Carpenter, and Paul-Christian
  B{\"u}rkner.
\newblock Rank-normalization, folding, and localization: an improved
  $\widehat{R}$ for assessing convergence of {MCMC}.
\newblock \emph{Bayesian Analysis}, 16\penalty0 (2):\penalty0 667--718, 2021.

\bibitem[Wadsworth et~al.(2017)Wadsworth, Tawn, Davison, and
  Elton]{wadsworth2017}
Jennifer~L. Wadsworth, Jonathan~A. Tawn, Anthony~C. Davison, and Daniel~M.
  Elton.
\newblock Modelling across extremal dependence classes.
\newblock \emph{Journal of the Royal Statistical Society, Series B},
  79:\penalty0 149--175, 2017.

\bibitem[Worms and Worms(2014)]{worms2014}
Julien Worms and Rym Worms.
\newblock New estimators of the extreme value index under random right
  censoring, for heavy-tailed distributions.
\newblock \emph{Extremes}, 17:\penalty0 337--358, 2014.

\end{thebibliography}

\clearpage
\setcounter{section}{0}
\setcounter{table}{0}
\setcounter{figure}{0}
\setcounter{equation}{0}
\renewcommand{\thesection}{S\arabic{section}}
\renewcommand{\thetable}{S\arabic{table}}
\renewcommand{\thefigure}{S\arabic{figure}}
\renewcommand{\theequation}{S\arabic{equation}}
\renewcommand{\theHsection}{S\arabic{section}}
\renewcommand{\theHtable}{S\arabic{table}}
\renewcommand{\theHfigure}{S\arabic{figure}}
\renewcommand{\theHequation}{S\arabic{equation}}

\pdfbookmark[1]{Supplement}{supplement}
\begin{center}
{\Large\bfseries Supplement to ``How much can a human being lift?''}
\end{center}
\bigskip

\section{The quadrature rules}
\label{sec:s-quad}

The likelihood \eqref{eq:lik} of the main text is computed by quadrature in two
layers. Dropping the superscript $l$, each athlete-day-lift requires
\[
I_i(v) = \int_0^{\infty} q_i(z; v)\, dz, \qquad
q_i(z; v) = \ell_i(z)\, g_i(z)\, \kappa_{\rho_a}[\Phi^{-1}\{G_i(z)\}, v],
\]
and each athlete $a$ requires $\int \prod_{i \in a} I_i(v)\, \phi(v)\, dv$.
This section gives the rule for each integral and checks that the
quadrature error is negligible.

\subsection{The integral over capability}
\label{sec:s-quad-z}

\emph{Support.} Let $\underline{w}_i$ be the lightest made weight of
athlete-day-lift $i$ and $\overline{w}_i$ the heaviest missed weight, and
for each declared attempt $j \in \mathcal{J}_i$ let $m_j$, $s_j^-$ and
$s_j^+$ be the mode and the two scales of its margin. The integrand is
negligible outside $[a_i, b_i]$, where
\begin{align*}
a_i &= \max\Bigl\{ \underline{w}_i (1 - 12\tau),\;
w_{i1}\, e^{m_1 - 8 s_1^-} \Bigr\},\\
b_i &= \min\Bigl\{ \overline{w}_i (1 + 12\tau),\;
\min_{j \in \mathcal{J}_i} w_{ij}\, e^{m_j + 8 s_j^+} \Bigr\}.
\end{align*}
The first terms come from the attempt model, since a lift at a weight
twelve attempt scales away from capability has a success or failure
probability of $F(-12) \approx 6 \times 10^{-6}$. The other terms come
from the margin densities, eight scales on either side of the mode. Only
the opener bounds capability from below. The density of a raise is divided
by its truncation probability, and for the smallest allowed raise, where
the declared weight sits at the truncation point, this ratio does not
decay below the mode. Without a made lift, the first term of $a_i$ is
replaced by $\mu(x_i) + \omega(x_i) \min_k (m_k - 8 s_k)$, eight standard
deviations below the lowest component, and without a miss the first term
of $b_i$ is dropped.

\emph{Breakpoints.} On $[a_i, b_i]$ the integrand is smooth except at the
threshold $z_u = \mu(x_i) + \omega(x_i) u$, where the capability density
may jump, and at the modes $z_j = w_{ij} e^{m_j}$, where a margin density
switches from one scale to the other and its derivative jumps. Some of the
scales are small, down to their lower bound of $0.3$ per cent of the bar,
so these kinks are sharp on the scale of the interval, and a margin with a
small lower scale rises steeply just below its mode.

\emph{Rule.} The interval is cut at $z_u$, at every $z_j$ inside it, and at
$z_j^- = w_{ij} e^{m_j - 4 s_j^-}$ for every margin whose lower side is
narrow, $s_j^- w_{ij} < 2h$. Each piece $[c, d]$ is integrated by
Gauss--Legendre quadrature with
\[
n = \begin{cases}
24, & d - c > 10, \\[2pt]
\min\{16, \max(4, \lceil (d - c)/h_{cd} \rceil)\}, & \text{otherwise},
\end{cases}
\]
\[
h = \max\Bigl\{\min\Bigl(0.7 \min_{j \in \mathcal{J}_i} s_j^+ w_{ij},\; 1\Bigr),\; 0.25\Bigr\},
\]
with lengths in kilograms, where $h_{cd} = h$ except on a piece
$[z_j^-, z_j]$, where it is the smaller of $h$ and
$\max(0.7 s_j^- w_{ij}, 0.25)$. A narrow piece thus gets a node spacing
below the narrowest margin scale it contains. A rule that splits
$[a_i, b_i]$ only at the threshold, with twenty-four nodes a side, misses
the kinks. For the men's snatch its total log likelihood is in error by
about forty units, and an optimizer finds spurious maxima where the narrow
densities fall between nodes.

\subsection{The integral over the athlete factor}

For athlete $a$ with $n_a$ competitions, let $\bar c_i$ be the mean of
$\Phi^{-1}\{G_i(Z_i)\}$ under the normalized integrand $q_i(z; 0)$ of each
competition, the capability on the normal scale as it would be estimated
with $\rho_a = 0$. Treating these as observations of
$C_i = \sqrt{\rho_a}\, V_a + \sqrt{1 - \rho_a}\, \epsilon_i$ gives the
approximate posterior mean and standard deviation of the factor,
\[
\hat v_a = \frac{\sqrt{\rho_a}}{1 - \rho_a}\,
\frac{\sum_{i \in a} \bar c_i}{1 + n_a \rho_a/(1 - \rho_a)},
\qquad
\hat s_a = \Bigl\{ 1 + \frac{n_a \rho_a}{1 - \rho_a} \Bigr\}^{-1/2}.
\]
The integral over $V_a$ uses twelve-point Gauss--Hermite nodes $x_k$ and
weights $\omega_k$ centred and scaled at these values,
\[
\int \prod_{i \in a} I_i(v)\, \phi(v)\, dv \approx
\sum_{k=1}^{12} \omega_k\, e^{x_k^2}\, \sqrt{2}\, \hat s_a\,
\phi(v_k) \prod_{i \in a} I_i(v_k),
\qquad v_k = \hat v_a + \sqrt{2}\, \hat s_a x_k,
\]
with each $I_i(v_k)$ computed as in Section~\ref{sec:s-quad-z}. A single competition
needs no outer integral.

\subsection{The integral over the shared error}

For the shared self-assessment error of Section~\ref{sec:s-sens}, the
margins are taken relative to $Z_i e^{\eta_i}$ with
$\eta_i \sim N(0, s_\eta^2)$, and the likelihood of an athlete-day-lift
includes an integral over $\eta_i$ inside the integral over capability.
The integrand in $\eta_i$ is a product of two-piece normal densities and
truncation factors. It is cut at the points where a margin or a truncation
point crosses its mode, with extra breakpoints near the truncation points.
On each segment the Gaussian part is integrated in closed form and the
remaining smooth factor by Gauss--Legendre quadrature with $K_\eta$ nodes.

\subsection{Accuracy}

Table~\ref{tab:s-quad} compares each rule with a more accurate reference.
In every case the discrepancy is far below the differences in log
likelihood between the anchors (Table~\ref{tab:s-sens}) and between the
fits with and without the shared error (Table~\ref{tab:s-shared}), which
run to hundreds of units. In addition, the mixture density integrates to one and its survival
function matches the integral of the density to six decimal places, and
at $\rho = 0$ the copula likelihood of the second stage, computed on the
product of the two grids, reproduces the sum of the two marginal log
likelihoods to $10^{-13}$.

\begin{table}[t]
\centering
\caption{Accuracy of the quadrature rules: absolute difference in log
likelihood between each rule and its reference. The shared-error rows are
at the posterior modes of Table~\ref{tab:s-shared}.}
\label{tab:s-quad}
\footnotesize\setlength{\tabcolsep}{3.5pt}
\resizebox{\textwidth}{!}{%
\begin{tabular}{lllr}
\toprule
Integral & Rule & Reference & Largest difference \\
\midrule
Capability & Rule of S1.1 & Grid of $400{,}001$ points & $10^{-3}$ per athlete-day-lift \\
Athlete factor & $12$ Gauss--Hermite nodes & $20$ nodes & $3 \times 10^{-3}$ per athlete \\
Athlete factor & $12$ Gauss--Hermite nodes & Grid of $4{,}001$ points & $3 \times 10^{-3}$ per athlete \\
Shared error & $K_\eta = 4$ & Grid of $200{,}001$ points & $5 \times 10^{-4}$ per athlete-day-lift \\
Shared error & $K_\eta = 4$ & $K_\eta = 64$ & 2 in total \\
\bottomrule
\end{tabular}}
\end{table}

\section{The anchor}
\label{sec:s-ridge}

\subsection{The simplified model along the anchor}

Table~\ref{tab:s-ridge} gives the fits behind Figure~\ref{fig:ridge} of the main text:
the simplified model of Section~\ref{sec:results-tau} of the main text for the men's snatch,
without the athlete factor, with one lower margin scale shared across the
three situations and with the normal-mixture body and estimated threshold
of the final model, with the relative attempt scale
$\tau$ fixed at the listed values and every other parameter re-estimated,
and the fit with $\tau$ free. Each fit is a quasi-Newton maximization of
the log posterior on a bounded reparametrization. Along a ridge a single
start can stop at a local mode, so each fit at fixed $\tau$ is started
twice, from the posterior mode of the final model and from the fit at the
neighbouring value of $\tau$, outwards from $\tau = 0.02$, and the better
of the two is kept. The two starts end within a tenth of a unit of log
posterior of each other at every value of $\tau$. The fit with $\tau$ free
starts from the kept fit at $\tau = 0.08$.

\begin{table}[t]
\centering
\caption{The simplified model for the men's snatch at fixed values of the
relative attempt scale: log likelihood relative to the fit at $\tau = 0.02$,
modes of the margin for openers and for raises after a success in per cent
of the bar, attempt-order shifts, tail shape and threshold, and the
one-in-a-thousand level and endpoint of the capability distribution at
$110$ kilograms of bodyweight in 2026.}
\label{tab:s-ridge}
\begin{tabular}{rrrrrrrrrr}
\toprule
$\tau$ & $\Delta\ell$ & $\hat m_1$ & $\hat m_2$ & $\hat\delta_2$ & $\hat\delta_3$ & $\hat\xi$ & $\hat u$ & $q_{0.999}$ & Endpoint \\
\midrule
0.010 & $-$736 & 2.9 & 1.2 & 1.55 & 2.42 & $-$0.32 & 1.23 & 197.8 & 202.6 \\
0.015 & $-$243 & 3.6 & 1.5 & 0.95 & 1.29 & $-$0.33 & 1.26 & 198.1 & 202.7 \\
0.020 & 0 & 4.5 & 1.8 & 0.64 & 0.71 & $-$0.34 & 1.32 & 198.5 & 202.8 \\
0.030 & 350 & 6.0 & 2.6 & 0.26 & 0.06 & $-$0.35 & 1.45 & 200.1 & 203.9 \\
0.050 & 565 & 8.3 & 4.8 & $-$0.10 & $-$0.52 & $-$0.37 & 1.29 & 204.1 & 207.5 \\
0.080 & 639 & 11.8 & 8.3 & $-$0.31 & $-$0.86 & $-$0.38 & 1.27 & 211.2 & 214.2 \\
free (0.120) & 656 & 16.6 & 13.1 & $-$0.42 & $-$1.06 & $-$0.39 & 1.25 & 221.3 & 224.3 \\
\bottomrule
\end{tabular}
\end{table}

\subsection{A simulation study}
\label{sec:s-sim}

The simulation study checks, on data with a known truth, that the final
model recovers its parameters at the correct anchor, and what a wrong
anchor does to them. Attempts are simulated from the final model at its
posterior mode at $\tau = 0.02$ for the men's snatch and the women's clean
and jerk, on the design of the data. Each athlete-day-lift keeps its
athlete, bodyweight, year and number of attempts. Capability is drawn with
the athlete factor and the opener from its margin density. After a success
the next weight is a raise, with the margin truncated as in Section~\ref{sec:method-decl} of
the main text. After a miss the weight is repeated with the observed share
of repeats and otherwise raised. Declared weights are rounded to whole
kilograms. Ten data sets are simulated for each combination and refitted by
the posterior mode at $\tau = 0.02$, and the first five are also refitted at
$\tau = 0.01$ and $0.03$. Every fit starts from the true values of the
regression, body, attempt and declaration parameters, with the tail
neutral, shape $0$ and threshold $1.5$, the centre of its prior.

Table~\ref{tab:s-sim} gives the results. At the correct anchor the
estimates centre on the truth in the men's snatch. There the tail shape has a
standard deviation of $0.04$ across replicates, the one-in-a-thousand level
a bias of $-0.3$ kilograms and a standard deviation of $0.6$, and the
endpoint a standard deviation of three kilograms. The women's clean and
jerk has a weaker tail. Its shape varies twice as much and is pulled
towards zero, by about half a standard deviation on average, one replicate in ten
is fitted with an unbounded tail, and the endpoint ranges from $178$ to
$221$ kilograms in the other nine, while the one-in-a-thousand level has a
standard deviation of $1.9$ kilograms. This matches the wide posterior of that ceiling in Table~\ref{tab:ceiling} of
the main text. A wrong anchor acts as Proposition~\ref{prop:ridge} describes. At
$\tau = 0.01$ the attempt-order shifts rise by about $1.1$ and $1.7$ and the
margin modes fall by $1.0$ to $1.6$ points of the bar, and at
$\tau = 0.03$ they move the other way. On average the tail shape and the
athlete correlation barely move, and the one-in-a-thousand level and the
endpoint move by at most $1.8$ kilograms. The real
data behave in the same way (Table~\ref{tab:s-sens}).

\begin{table}[t]
\centering
\caption{Simulation study. True values (the posterior mode at
$\tau = 0.02$), mean and standard deviation of the estimates over ten
simulated data sets refitted at $\tau = 0.02$, and mean change when the
first five are refitted at $\tau = 0.01$ and $0.03$, against the fit at
$\tau = 0.02$ to the same data. Quantiles and endpoints at the reference
bodyweight in 2026. $^{*}$Median, one replicate has an unbounded tail.
$^{\dagger}$Over the replicates with a bounded tail in both fits.}
\label{tab:s-sim}
\footnotesize
\begin{tabular}{lrrrr}
\toprule
 & Truth & Fit at $\tau = 0.02$ & \multicolumn{2}{c}{Change at} \\
\cmidrule(lr){4-5}
 & & & $\tau = 0.01$ & $\tau = 0.03$ \\
\midrule
\multicolumn{5}{l}{\emph{Men's snatch}} \\
Tail shape $\xi$ & $-$0.308 & $-$0.293 (0.040) & 0.041 & $-$0.001 \\
Threshold $u$ & 1.68 & 2.00 (0.28) & 0.00 & 0.01 \\
Athlete correlation $\rho_a$ & 0.906 & 0.905 (0.003) & $-$0.004 & 0.002 \\
Shift $\delta_2$ & 0.66 & 0.66 (0.03) & 1.11 & $-$0.45 \\
Shift $\delta_3$ & 0.75 & 0.75 (0.05) & 1.80 & $-$0.68 \\
Opener mode $m_1$ (\%) & 3.26 & 3.24 (0.11) & $-$1.59 & 1.44 \\
Raise mode $m_2$ (\%) & 1.50 & 1.53 (0.09) & $-$0.99 & 1.14 \\
$q_{0.999}$ (kg) & 199.2 & 198.9 (0.6) & $-$0.5 & 1.8 \\
Endpoint (kg) & 206.0 & 206.5 (2.9) & 1.3 & 1.8 \\
\multicolumn{5}{l}{\emph{Women's clean and jerk}} \\
Tail shape $\xi$ & $-$0.177 & $-$0.139 (0.079) & 0.031 & $-$0.002 \\
Threshold $u$ & 1.59 & 1.71 (0.12) & 0.11 & $-$0.02 \\
Athlete correlation $\rho_a$ & 0.928 & 0.928 (0.002) & $-$0.006 & 0.003 \\
Shift $\delta_2$ & 0.43 & 0.41 (0.03) & 1.06 & $-$0.45 \\
Shift $\delta_3$ & 0.24 & 0.23 (0.04) & 1.68 & $-$0.67 \\
Opener mode $m_1$ (\%) & 0.73 & 0.73 (0.08) & $-$1.10 & 1.31 \\
Raise mode $m_2$ (\%) & 0.13 & 0.18 (0.21) & $-$1.20 & 1.61 \\
$q_{0.999}$ (kg) & 167.6 & 167.8 (1.9) & $-$0.6 & 1.6 \\
Endpoint (kg) & 186.9 & 190.8$^{*}$ & 0.5$^{\dagger}$ & 1.5$^{\dagger}$ \\
\bottomrule
\end{tabular}
\end{table}

\section{The tail without a body}
\label{sec:s-direct}

Table~\ref{tab:s-direct} gives the check without a body used in the last
row of Table~\ref{tab:ladder} of the main text. The threshold $u(x)$ is the $q$-quantile
of the best made lift of an athlete-day, estimated by linear quantile
regression on the design of the main text (intercept, standardized log
bodyweight, its orthogonal square and centred year). The exceedances
$y - u(x)$ follow a generalized Pareto distribution with a log scale linear
in standardized log bodyweight and a constant shape, fitted by maximum
likelihood treating the athlete-days as independent, and the intervals come
from $1{,}000$ bootstrap samples that resample athletes with all their
competitions, which carries the dependence between the competitions of the
same athlete into the
intervals. The endpoint at covariate $x$ is $u(x) + \sigma(x)/(-\xi)$. The
shape is negative at both thresholds in all four combinations. Between the
two thresholds the endpoint moves by at most four kilograms for men and in
the women's snatch, and by seven in the women's clean and jerk, whose
shape falls from $-0.21$ to $-0.31$. At the higher threshold the shape is
estimated from about five hundred exceedances, and its bootstrap
interval widens to about $0.3$. At the lightest women's bodyweights the
endpoint falls at or below the record (at $59$ kilograms, $116$ against a
snatch record of $117$ at the lower threshold and $149$ against a clean and
jerk record of $149$ at the higher), a sign that a scale linear in log bodyweight is
too rigid at the edge of the range.

\begin{table}[t]
\centering
\caption{Generalized Pareto fits without a body: threshold quantile, number
of exceedances, tail shape with athlete-bootstrap $95$ per cent interval,
endpoint at the reference bodyweight ($110$ kilograms for men, $86$ for
women) in 2026, and the local record of the current era.}
\label{tab:s-direct}
\begin{tabular}{llrrrrr}
\toprule
Sex & Lift & $q$ & Exceedances & $\hat\xi$ & Endpoint & Record \\
\midrule
Men & Snatch & 0.90 & 1{,}007 & $-$0.24 ($-$0.34, $-$0.16) & 212 & 200 \\
 &  & 0.95 & 502 & $-$0.21 ($-$0.39, $-$0.08) & 212 &  \\
Men & Clean and jerk & 0.90 & 965 & $-$0.22 ($-$0.31, $-$0.15) & 255 & 242 \\
 &  & 0.95 & 479 & $-$0.20 ($-$0.34, $-$0.08) & 259 &  \\
Women & Snatch & 0.90 & 933 & $-$0.32 ($-$0.43, $-$0.20) & 140 & 127 \\
 &  & 0.95 & 468 & $-$0.35 ($-$0.51, $-$0.23) & 140 &  \\
Women & Clean and jerk & 0.90 & 917 & $-$0.21 ($-$0.33, $-$0.08) & 187 & 161 \\
 &  & 0.95 & 456 & $-$0.31 ($-$0.46, $-$0.19) & 180 &  \\
\bottomrule
\end{tabular}
\end{table}

\section{Fixed and estimated threshold}
\label{sec:s-splice}

Table~\ref{tab:s-splice} fits the record book, the best made lifts with the
athlete-days exchangeable, with the threshold of the mixture body held at
the $0.85$, $0.90$ and $0.95$ quantiles of the standardized error and every
other parameter re-estimated. The fits are by maximum likelihood with the
density of each best lift, whereas the record-book rows of Table~\ref{tab:ladder} of the
main text are posterior summaries that use the probability of the
one-kilogram interval around each lift and the priors of the full model.
The ceiling moves with the threshold, by up
to four kilograms in the men's lifts and the women's snatch and by ten in
the women's clean and jerk. A threshold fixed in advance carries an uncertainty that intervals at that threshold do not show. In the
latent model the threshold is a parameter, and its posterior carries this uncertainty into every tail quantity.

\begin{table}[t]
\centering
\caption{Tail shape and ceiling at the reference bodyweight in 2026 for the
record book, the best made lifts with the athlete-days exchangeable, with
the threshold of the mixture body held at three quantiles of the
standardized error.}
\label{tab:s-splice}
\begin{tabular}{llrrrr}
\toprule
 & & Men, snatch & Men, c\&j & Women, snatch & Women, c\&j \\
\midrule
Threshold at 0.85 & $\hat\xi$ & $-$0.24 & $-$0.23 & $-$0.23 & $-$0.20 \\
 & ceiling & 212 & 255 & 144 & 188 \\
Threshold at 0.90 & $\hat\xi$ & $-$0.24 & $-$0.23 & $-$0.28 & $-$0.22 \\
 & ceiling & 212 & 255 & 142 & 185 \\
Threshold at 0.95 & $\hat\xi$ & $-$0.33 & $-$0.26 & $-$0.29 & $-$0.33 \\
 & ceiling & 208 & 253 & 141 & 178 \\
\bottomrule
\end{tabular}
\end{table}

The threshold is harder to place in the record-book fits of Table~\ref{tab:ladder} of the
main text than in the full model. In the record-book fit of the women's
snatch with the athlete-days exchangeable, the posterior of the threshold
is wide and the chains explore it slowly, with an effective sample size of
$82$ for the threshold against $128$ for the shape
(Table~\ref{tab:s-conv}). At the higher thresholds the shape
is estimated from a few hundred exceedances and its posterior spreads
towards zero, which is why that row of Table~\ref{tab:ladder} gives a finite ceiling with
probability $0.84$ only. In the full model the women's snatch threshold is
$1.47$ $(0.94, 2.48)$, an interval that spans the two regions of its
posterior (Section~\ref{sec:s-conv}).

\section{Model checks and sensitivity analyses}
\label{sec:s-sens}
\label{sec:s-ppc}

Table~\ref{tab:s-ppc} compares the observed success rate of each attempt
with the rate predicted by the final model from everything observed before
its outcome, the earlier weights and outcomes and the weight just declared,
with the parameters at their posterior mode and the athlete factor switched
off, so that the prediction uses the day alone. The rates by attempt agree
to within $0.035$ in every combination, and the rates by size of raise are in
Figure~\ref{fig:jump} of the main text.

\begin{table}[t]
\centering
\caption{Observed and predicted success rates by sex, lift and attempt
number, the prediction for each attempt computed at the posterior mode from
the weights and outcomes observed before it. For the women's snatch the mode
is that of the region with the larger posterior mass.}
\label{tab:s-ppc}
\begin{tabular}{llrrr}
\toprule
Sex & Lift & Attempt & Observed & Predicted \\
\midrule
Men & Snatch & 1 & 0.771 & 0.761 \\
Men & Snatch & 2 & 0.633 & 0.619 \\
Men & Snatch & 3 & 0.437 & 0.434 \\
Men & Clean and jerk & 1 & 0.795 & 0.784 \\
Men & Clean and jerk & 2 & 0.601 & 0.588 \\
Men & Clean and jerk & 3 & 0.348 & 0.347 \\
Women & Snatch & 1 & 0.772 & 0.756 \\
Women & Snatch & 2 & 0.651 & 0.623 \\
Women & Snatch & 3 & 0.466 & 0.455 \\
Women & Clean and jerk & 1 & 0.826 & 0.818 \\
Women & Clean and jerk & 2 & 0.632 & 0.598 \\
Women & Clean and jerk & 3 & 0.372 & 0.362 \\
\bottomrule
\end{tabular}
\end{table}

Table~\ref{tab:s-ppcrec} checks the record book itself. For $200$ draws
from the posterior, complete competitions are simulated for every
athlete-day-lift of the data as in Section~\ref{sec:s-sim}, and the local
record of the current era is recomputed at each bodyweight of Table~\ref{tab:quant} of
the main text. The observed record lies within the central $90$ per cent of
the simulated records at $36$ of the $40$ combinations. The men's snatch
record at $61$ kilograms lies one kilogram below that range, and the
records of the heaviest women, in the snatch at $100$ kilograms and in the
clean and jerk at $100$ and $120$, lie two to five kilograms above it. On
the scale of capability, $(\text{best} - \mu)/\omega$, the $0.999$ quantile
of the best made lifts is reproduced in three of the four combinations,
but the $0.99$ quantile is higher than in at least $95$ per cent of the
simulated data sets in all four. Among the top one per cent of best lifts,
$0.76$ to $0.84$ come from a day that ended with a made attempt, against
$0.54$ to $0.64$ in the simulated competitions, and at most one simulated
data set in two hundred reaches the observed share. Under the model the outcome of an attempt
depends on its margin and not on the level of capability, so the best days
of the strongest lifters end with a miss as often as anyone's. In the data
they end with a make more often. The model therefore reads some of these made final
attempts as luck, which pulls the inferred capability of the strongest days
down.

\begin{table}[t]
\centering
\caption{Predictive check of the record book. Local record of the current
era (the heaviest lift made since 2019 within five kilograms of the
bodyweight) and the median and $90$ per cent interval of the same record in
competitions simulated from $200$ posterior draws, in kilograms.}
\label{tab:s-ppcrec}
\footnotesize
\begin{tabular}{rrrrr}
\toprule
Bodyweight & \multicolumn{2}{c}{Snatch} & \multicolumn{2}{c}{Clean and jerk} \\
\cmidrule(lr){2-3}\cmidrule(lr){4-5}
 & Record & Simulated & Record & Simulated \\
\midrule
\multicolumn{5}{l}{\emph{Men}} \\
61 & 146 & 151 (147, 157) & 183 & 184 (179, 190) \\
67 & 160 & 161 (156, 168) & 197 & 195 (190, 202) \\
73 & 169 & 168 (163, 174) & 209 & 203 (198, 211) \\
81 & 182 & 177 (172, 182) & 211 & 214 (209, 221) \\
89 & 188 & 187 (182, 193) & 229 & 225 (221, 233) \\
96 & 188 & 192 (187, 199) & 233 & 231 (226, 238) \\
102 & 190 & 195 (190, 203) & 233 & 236 (230, 243) \\
110 & 200 & 200 (194, 207) & 242 & 241 (235, 248) \\
130 & 204 & 204 (195, 214) & 250 & 248 (238, 257) \\
150 & 217 & 211 (201, 221) & 253 & 255 (243, 265) \\
\multicolumn{5}{l}{\emph{Women}} \\
49 & 99 & 103 (99, 109) & 126 & 129 (123, 138) \\
55 & 111 & 111 (107, 117) & 141 & 139 (134, 146) \\
59 & 117 & 117 (113, 122) & 149 & 145 (141, 154) \\
64 & 120 & 119 (115, 126) & 151 & 149 (143, 156) \\
71 & 125 & 124 (120, 132) & 156 & 156 (150, 165) \\
76 & 125 & 128 (123, 135) & 161 & 160 (152, 170) \\
81 & 127 & 130 (125, 136) & 161 & 163 (155, 171) \\
86 & 127 & 130 (125, 137) & 161 & 163 (156, 172) \\
100 & 145 & 131 (122, 140) & 178 & 164 (155, 176) \\
120 & 140 & 134 (125, 146) & 185 & 170 (159, 182) \\
\bottomrule
\end{tabular}
\end{table}

Table~\ref{tab:s-miss} gives the margin of a raise after a missed attempt,
which is left out of Table~\ref{tab:attempt} of the main text. Such a raise is truncated
above at $\log\{Z/(w_{j-1} + 1)\}$, and after a miss this point is usually
near or below zero, so the observed margins come from the part of the
two-piece normal below its mode. The data determine only the shape of the
density below the truncation point, which the mode and the two scales
describe together, so the mode is not a typical margin and the three
parameters are barely separable. In three of the four combinations the mode
lies far above the truncation point, against the upper end of its support
at $30$ per cent (Section~\ref{sec:method-prior} of the main text), and the upper scale is barely
determined, as its intervals show. In the women's snatch the mode lies just
below zero, with the lower scale at the lower end of its support. About
three athletes in four repeat the weight after a miss, and the choice
between repeating and raising is not modelled.

\begin{table}[t]
\centering
\caption{Posterior medians and $95$ per cent intervals for the margin of a
raise after a missed attempt at the reference anchor $\tau = 0.02$, in per
cent of the bar.}
\label{tab:s-miss}
\footnotesize\setlength{\tabcolsep}{3.5pt}
\begin{tabular}{llrrrr}
\toprule
 & & Men, snatch & Men, c\&j & Women, snatch & Women, c\&j \\
\midrule
Raise after a miss & mode & 24.4 (15.5, 29.7) & 28.7 (24.4, 29.9) & $-$0.5 ($-$0.7, $-$0.3) & 27.2 (20.5, 29.9) \\
 & upper scale & 3.1 (0.5, 20.8) & 2.9 (0.5, 21.5) & 1.2 (1.0, 1.6) & 2.9 (0.5, 19.9) \\
 & lower scale & 4.2 (3.5, 4.7) & 5.0 (4.6, 5.2) & 0.3 (0.3, 0.3) & 4.8 (4.2, 5.1) \\
\bottomrule
\end{tabular}
\end{table}

Table~\ref{tab:s-sens} reports posterior modes of the final model for the
four combinations of sex and lift at the three anchors, for the women's
snatch at both modes of Section~\ref{sec:s-conv}, and with $\tau$
left free for the men's snatch. Between $\tau = 0.01$ and $0.03$ the tail
shape moves by at most $0.05$, the one-in-a-thousand level by at most
$4.5$ kilograms and the endpoint by at most five, except at the
higher-threshold mode of the women's snatch, where the shape moves from
$-0.11$ to $-0.05$ at $\tau = 0.01$ and the endpoint from $152$ to $185$
kilograms. At the other mode, which holds most of the mass, the endpoint
lies between $142$ and $146$ kilograms at the three anchors. The
attempt-order shifts take up most of the change and fall in every
combination as $\tau$ grows. The mode of the opener margin rises between
$\tau = 0.01$ and $0.02$, roughly doubling in three of the four
combinations. Between $0.02$ and $0.03$ it keeps rising in the clean and
jerk, changes little in the men's snatch and falls in the women's snatch,
where the lower scale of the opener margin shrinks with it along a
direction that the data barely resolve. For that reason each anchor fit of
the women's snatch was also started from the margins of the fit at the
other mode, and the table reports the higher of the two. With $\tau$ free
the fit moves far along the direction of Proposition~\ref{prop:ridge}, with the margins,
the upper quantiles and the endpoint moving together. The optimizer ends at
$\tau = 0.144$ with the tail shape at the upper limit of its support, the
last row of the table.

\begin{table}[t]
\centering
\caption{Posterior modes of the final model at the three anchors, for the
women's snatch at both modes of its posterior
(Section~\ref{sec:s-conv}): log
likelihood, mode of the opener margin in per cent of the bar,
attempt-order shifts, tail shape, and the one-in-a-thousand level and
endpoint of the capability distribution at the reference bodyweight
($110$ kilograms for men, $86$ for women) in 2026, with the mean difference
between capability and the best made lift in kilograms.}
\label{tab:s-sens}
\resizebox{\textwidth}{!}{%
\begin{tabular}{lllrrrrrrrr}
\toprule
Sex & Lift & Fit & $\ell$ & $\hat m_1$ & $\hat\delta_2$ & $\hat\delta_3$ & $\hat\xi$ & $q_{0.999}$ & Endpoint & Gap \\
\midrule
Men & Snatch & $\tau = 0.01$ & $-$90{,}671.7 & 1.7 & 1.57 & 2.46 & $-$0.29 & 198.8 & 206.6 & $-$1.4 \\
Men & Snatch & $\tau = 0.02$ & $-$90{,}025.4 & 3.3 & 0.66 & 0.75 & $-$0.31 & 199.2 & 206.0 & $-$0.4 \\
Men & Snatch & $\tau = 0.03$ & $-$89{,}619.5 & 3.2 & 0.34 & 0.14 & $-$0.30 & 203.3 & 210.8 & 0.7 \\
Men & Clean and jerk & $\tau = 0.01$ & $-$92{,}718.5 & 1.9 & 1.21 & 1.70 & $-$0.36 & 241.4 & 247.1 & $-$0.6 \\
Men & Clean and jerk & $\tau = 0.02$ & $-$92{,}191.5 & 3.6 & 0.37 & 0.15 & $-$0.41 & 240.8 & 244.7 & 0.7 \\
Men & Clean and jerk & $\tau = 0.03$ & $-$91{,}913.9 & 4.8 & 0.07 & $-$0.42 & $-$0.41 & 244.5 & 248.5 & 2.2 \\
Women & Snatch & $\tau = 0.01$ & $-$73{,}599.2 & 0.9 & 2.02 & 3.13 & $-$0.05 & 131.1 & 184.8 & $-$1.1 \\
Women & Snatch & $\tau = 0.02$ & $-$72{,}594.9 & 1.8 & 0.83 & 1.07 & $-$0.11 & 131.5 & 152.2 & $-$0.4 \\
Women & Snatch & $\tau = 0.03$ & $-$72{,}337.3 & 0.6 & 0.42 & 0.35 & $-$0.10 & 132.6 & 156.5 & 0.3 \\
Women & Snatch, second mode & $\tau = 0.01$ & $-$73{,}592.9 & 0.9 & 2.02 & 3.13 & $-$0.22 & 131.7 & 143.4 & $-$1.1 \\
Women & Snatch, second mode & $\tau = 0.02$ & $-$72{,}588.4 & 1.8 & 0.83 & 1.07 & $-$0.24 & 132.1 & 141.9 & $-$0.4 \\
Women & Snatch, second mode & $\tau = 0.03$ & $-$72{,}326.5 & 0.6 & 0.42 & 0.35 & $-$0.20 & 133.3 & 145.9 & 0.3 \\
Women & Clean and jerk & $\tau = 0.01$ & $-$77{,}705.5 & $-$0.5 & 1.43 & 1.95 & $-$0.16 & 167.8 & 191.1 & $-$0.4 \\
Women & Clean and jerk & $\tau = 0.02$ & $-$77{,}333.0 & 0.7 & 0.43 & 0.24 & $-$0.18 & 167.6 & 186.9 & 0.6 \\
Women & Clean and jerk & $\tau = 0.03$ & $-$77{,}268.4 & 2.1 & 0.01 & $-$0.43 & $-$0.18 & 169.0 & 187.1 & 1.9 \\
Men & Snatch & $\tau$ free ($\hat\tau = 0.144$) & $-$88{,}676.5 & 17.1 & $-$0.43 & $-$1.09 & 0.40 & 236.4 & $\infty$ & 20.8 \\
\bottomrule
\end{tabular}}
\end{table}

Table~\ref{tab:s-prior} checks the priors of the tail. All posterior
draws of each combination are reweighted by the ratio of an
alternative prior to the original one, which gives the posterior under the
alternative without new chains. The alternatives are a wider prior for the
tail shape, $\xi_l \sim N(0, 0.5^2)$, a wider prior for the threshold,
$u_l \sim N(1.5, 2^2)$ on the same support, and a threshold bounded below
$3$. The effective sample size of the weights is at least $75$ per cent of
the number of draws. The median ceiling at the reference bodyweight moves by
at most $1.1$ kilograms and the one-in-a-thousand level by at most $0.1$.
The probability of a finite ceiling moves by at most $0.02$, except in the
women's clean and jerk under the wider prior for the threshold, where it
falls from $0.94$ to $0.89$, as that prior favours higher thresholds with
fewer lifts above them. The upper limit of the interval for the women's
ceilings is set by the share of draws with an unbounded tail, and under
the wider prior for the threshold it is infinite in both lifts.

\begin{table}[t]
\centering
\caption{Prior sensitivity of the tail by reweighting of the posterior
draws: effective sample size of the weights, posterior medians of the tail
shape, the threshold and the one-in-a-thousand level, posterior probability
of a finite ceiling, and posterior median and $95$ per cent interval of
the ceiling, infinite draws included, at the reference
bodyweight ($110$ kilograms for men, $86$ for
women) in 2026.}
\label{tab:s-prior}
\resizebox{\textwidth}{!}{%
\begin{tabular}{lllrrrrrr}
\toprule
Sex & Lift & Prior & ESS & $\xi$ & $u$ & $q_{0.999}$ & $P(\text{finite})$ & Ceiling \\
\midrule
Men & Snatch & Reference & 18{,}000 & $-$0.31 & 1.75 & 199.3 & 1.000 & 206.1 (203.3, 209.9) \\
 &  & $\xi_l \sim N(0, 0.5^2)$ & 17{,}928 & $-$0.31 & 1.75 & 199.3 & 1.000 & 206.0 (203.2, 209.8) \\
 &  & $u_l \sim N(1.5, 2^2)$ & 17{,}671 & $-$0.31 & 1.80 & 199.3 & 1.000 & 206.1 (203.3, 209.9) \\
 &  & $u_l < 3$ & 17{,}997 & $-$0.31 & 1.75 & 199.3 & 1.000 & 206.1 (203.3, 209.9) \\
Men & Clean and jerk & Reference & 18{,}000 & $-$0.40 & 2.00 & 241.0 & 1.000 & 245.0 (243.0, 247.3) \\
 &  & $\xi_l \sim N(0, 0.5^2)$ & 17{,}938 & $-$0.41 & 2.00 & 240.9 & 1.000 & 244.9 (243.0, 247.2) \\
 &  & $u_l \sim N(1.5, 2^2)$ & 14{,}373 & $-$0.40 & 2.22 & 241.0 & 1.000 & 245.0 (243.1, 247.2) \\
 &  & $u_l < 3$ & 16{,}875 & $-$0.41 & 1.94 & 241.0 & 1.000 & 245.0 (243.0, 247.3) \\
Women & Snatch & Reference & 43{,}326 & $-$0.23 & 1.47 & 132.1 & 0.955 & 142.8 (139.0, 418.1) \\
 &  & $\xi_l \sim N(0, 0.5^2)$ & 43{,}079 & $-$0.23 & 1.46 & 132.1 & 0.960 & 142.6 (138.8, 241.5) \\
 &  & $u_l \sim N(1.5, 2^2)$ & 42{,}682 & $-$0.22 & 1.48 & 132.1 & 0.945 & 143.0 (139.0, $\infty$) \\
 &  & $u_l < 3$ & 43{,}314 & $-$0.23 & 1.47 & 132.1 & 0.955 & 142.8 (139.0, 418.1) \\
Women & Clean and jerk & Reference & 66{,}000 & $-$0.18 & 1.63 & 167.8 & 0.941 & 187.1 (178.3, $\infty$) \\
 &  & $\xi_l \sim N(0, 0.5^2)$ & 65{,}203 & $-$0.18 & 1.63 & 167.8 & 0.941 & 186.6 (178.0, $\infty$) \\
 &  & $u_l \sim N(1.5, 2^2)$ & 49{,}337 & $-$0.17 & 1.71 & 167.9 & 0.888 & 188.2 (178.3, $\infty$) \\
 &  & $u_l < 3$ & 63{,}523 & $-$0.18 & 1.62 & 167.7 & 0.957 & 186.9 (178.3, 359.3) \\
\bottomrule
\end{tabular}}
\end{table}

Table~\ref{tab:s-trend} adds a linear term in the centred year to the log
scale of capability in the record-book fit of the first row of Table~\ref{tab:ladder} of
the main text, at its posterior mode, whose values differ from the
posterior medians of that table. The term is negligible. Its coefficient is under $0.3$ per cent a year in absolute
value, the log likelihood improves by at most $1.6$ units, and the
one-in-a-thousand levels move by under half a kilogram. The endpoint of
the women's snatch moves by $7.5$ kilograms, because the shape of that fit
is weakly determined (Section~\ref{sec:s-splice}).

\begin{table}[t]
\centering
\caption{The record-book fit with and without a linear time term in the log
scale: improvement in log likelihood, coefficient of the centred year in
the log scale, and the tail shape, one-in-a-thousand level and endpoint at
the reference bodyweight in 2026 without and with the term.}
\label{tab:s-trend}
\resizebox{\textwidth}{!}{%
\begin{tabular}{llrrrrrrrr}
\toprule
 & & & & \multicolumn{2}{c}{$\hat\xi$} & \multicolumn{2}{c}{$q_{0.999}$} & \multicolumn{2}{c}{Endpoint} \\
\cmidrule(lr){5-6}\cmidrule(lr){7-8}\cmidrule(lr){9-10}
Sex & Lift & $\Delta\ell$ & $\hat\gamma_{l3}$ & without & with & without & with & without & with \\
\midrule
Men & Snatch & 0.5 & $-$0.0015 & $-$0.31 & $-$0.31 & 202.3 & 202.2 & 208.6 & 208.4 \\
Men & Clean and jerk & 1.2 & 0.0026 & $-$0.21 & $-$0.21 & 243.9 & 244.2 & 256.3 & 257.0 \\
Women & Snatch & 0.6 & 0.0008 & $-$0.14 & $-$0.09 & 133.8 & 133.8 & 148.4 & 155.9 \\
Women & Clean and jerk & 1.6 & $-$0.0004 & $-$0.26 & $-$0.28 & 171.7 & 171.3 & 183.5 & 181.6 \\
\bottomrule
\end{tabular}}
\end{table}

Two variants of the attempt model were also fitted at the reference
anchor. The first adds a shared self-assessment error to the final model.
The athlete's estimate of capability on the day is $Z_i e^{\eta_i}$, with
$\eta_i \sim N(0, s_\eta^2)$ common to the day's declarations and
$\log s_\eta \sim N(\log 0.02, 1)$ on $(0.002, 0.3)$. The
margins are taken relative to that estimate. Table~\ref{tab:s-shared}
compares the posterior modes with and without it for the four combinations
of sex and lift, for the women's snatch at both modes. The shared error is
between $1.7$ and $3.0$ per cent of
capability and improves the log likelihood by $95$ units in the women's
snatch and by $600$ to $900$ units in the other three combinations. It
takes up part of the attempt-order shifts, which fall by $0.03$ to $0.24$,
and leaves the athlete correlation within $0.01$ and the one-in-a-thousand
level at the reference bodyweight within $1.3$ kilograms. The tail shape
moves by at most $0.02$ and the endpoint by at most $2.4$ kilograms,
except at the higher-threshold mode of the women's snatch, where the shape
moves from $-0.11$ to $-0.04$ and the endpoint from $152$ to $192$
kilograms. That mode holds a fifth of the posterior mass
(Section~\ref{sec:s-conv}), and its endpoint responds to any change that
moves the shape towards zero, as it does at the anchors
(Table~\ref{tab:s-sens}). At the mode that holds the rest of the mass the
shape moves from $-0.24$ to $-0.22$ and the endpoint by $0.8$ kilograms. The shared
error does not change the success curve by
size of raise, because the size of a raise is the ratio of two
declarations made relative to the same estimate and so does not depend on
the shared error. Each likelihood evaluation costs about twenty times as
much as for the final model, which puts the chains out of reach, so the
shared error is left out of the final model. The second variant, fitted to
the men's snatch with the normal-mixture body and without the athlete
factor, puts a floor on the miss probability,
$P(S = 1 \mid Z, w) = (1 - \pi) F\{\log(Z/w)/\tau\}$. It improves the log
likelihood by $819$ units for one parameter, but does so by moving along
the direction of Proposition~\ref{prop:ridge}, with $\hat\pi = 0.32$, openers declared
$12$ per cent below capability, the order shifts at $-0.6$ and $-2.5$ and
the endpoint at $223$ kilograms, against $203$ without the floor. The
floor and the shift of the capability
scale are indistinguishable to the data, and the variant is not used.

\begin{table}[t]
\centering
\caption{The final model with and without the shared self-assessment
error, posterior modes at the reference anchor: improvement in log
likelihood, scale of the shared error, and the tail shape, the
one-in-a-thousand level and the endpoint at the reference bodyweight
($110$ kilograms for men, $86$ for women) in 2026, without and with the
shared error.}
\label{tab:s-shared}
\resizebox{\textwidth}{!}{%
\begin{tabular}{llrrrrrrrr}
\toprule
 & & & & \multicolumn{2}{c}{$\hat\xi$} & \multicolumn{2}{c}{$q_{0.999}$} & \multicolumn{2}{c}{Endpoint} \\
\cmidrule(lr){5-6}\cmidrule(lr){7-8}\cmidrule(lr){9-10}
Sex & Lift & $\Delta\ell$ & $\hat s_\eta$ & without & with & without & with & without & with \\
\midrule
Men & Snatch & 875 & 0.027 & $-$0.31 & $-$0.32 & 199.2 & 199.6 & 206.0 & 205.7 \\
Men & Clean and jerk & 849 & 0.027 & $-$0.41 & $-$0.40 & 240.8 & 241.4 & 244.7 & 245.3 \\
Women & Snatch & 95 & 0.017 & $-$0.11 & $-$0.04 & 131.5 & 131.1 & 152.2 & 191.6 \\
Women & Snatch, second mode & 95 & 0.017 & $-$0.24 & $-$0.22 & 132.1 & 132.0 & 141.9 & 142.7 \\
Women & Clean and jerk & 615 & 0.030 & $-$0.18 & $-$0.17 & 167.6 & 168.9 & 186.9 & 189.3 \\
\bottomrule
\end{tabular}}
\end{table}

A linear trend in time is kept in the location. Its coefficient is $0.32$
and $0.36$ kilograms a year for the men's two lifts and $0.60$ and $0.79$
for the women's (Table~\ref{tab:tail} of the main text).

\section{Convergence of the chains}
\label{sec:s-conv}

The final model was sampled, for each sex and lift, by four
differential-evolution Markov chains with sampling from the past and
snooker updates \citep{terbraak2008}. The chains work in the coordinates of
the main text with the threshold-stable tail scale
$\sigma_l^{*} = \sigma_l - \xi_l u_l$, which does not move with the
threshold \citep[Section 4.3.4]{coles2001}. The change of variables is
exact, since the target density carries its Jacobian. The sampler was
chosen because the threshold, the tail scale, the free mixture components
and the regression coefficients lie along curved ridges of the posterior,
on which a Gaussian random walk gave effective sample sizes of about twenty
in $17{,}000$ draws. Each chain ran for $120{,}000$ generations, the first
$30{,}000$ were discarded and every twentieth state of the rest was kept,
$4{,}500$ draws per chain, or $360{,}000$ generations and $16{,}500$ draws
for the two women's lifts. The shared archive of past states was extended
every ten generations. The initial archive and the starting states were
drawn around the posterior mode with four times the variance of its
Laplace approximation and with the threshold spread over half a scale unit
on either side. The record-book fits of Table~\ref{tab:ladder} of the main text were
sampled in the same way with $200{,}000$ generations, $50{,}000$ discarded
and every fiftieth state kept.

The posterior mode is the best of three optimizations. One starts from a fit to the best lifts and one from a high threshold with a positive shape. The third moves the first start along the ridge that links the scale of capability to the free mixture components. For the men's lifts and the women's clean and jerk the three
optimizations end at the same mode or within $1.6$ units of log posterior
of it. For the women's snatch they end in two regions. One has a threshold
of $2.2$ and a shape of $-0.11$. The other, $4.7$ units of log posterior
higher, has a narrower scale of capability, a threshold of $1.4$ and a
shape of $-0.24$. The replication package stores it as a start, because
the optimizer reaches it only in some runs. Four chains were run around each mode, for $360{,}000$ generations at the higher-threshold mode and $120{,}000$ at the other. Every retained draw
is assigned to the nearer of the two modes in the threshold, the log scale
and the shape, each divided by its standard deviation within a chain.
These chains cross between the regions rarely, so the share of each region
in their pooled draws would depend on where they started. A third set of eight chains therefore starts from the last states of these eight, with an archive of $3{,}200$ draws taken from both sets in equal numbers. Every second proposal is a jump of full length along the difference of two archive states, which can carry a chain from one region to the other. The
proposals are symmetric, so the set has the posterior as its stationary
distribution whatever the archive holds. After the first tenth of its
$40{,}000$ generations it crossed between the regions $41$ times, about once every $7{,}000$ generations per chain. The share of its draws in each region estimates the posterior mass of the region, $0.78$ for the region of the narrow mode, with a standard error of $0.04$ from ten batches of each chain.
The draws of each region from all sixteen chains are pooled with these
weights. The regions agree on the one-in-a-thousand level, $132.2$ against
$131.8$ kilograms at the reference bodyweight, but differ on the endpoint,
$142$ kilograms against $158$ with $21$ per cent of the draws at shapes at
or above zero. The checks at the mode in Section~\ref{sec:s-sens} are made
at both modes.

Table~\ref{tab:s-conv} gives the acceptance rates with the rank-normalized split $\widehat{R}$ and the bulk effective sample sizes of \citet{vehtari2021}, for the women's snatch within each region. In the men's fits and the
women's clean and jerk, $\widehat{R}$ is at most $1.03$ for the tail shape,
the athlete correlation, the attempt and declaration parameters, the time
trend and the derived quantities. It reaches $1.05$ only among the means,
scales and weights of the two free mixture components, which are
exchangeable and not what the paper reports. Within the two regions of the
women's snatch the tail shape, the endpoint and the one-in-a-thousand level
have $\widehat{R}$ at most $1.11$ and effective sample sizes of at least
$330$, while the parameters of the body reach $1.18$ and $1.33$ along the
ridge in the threshold and the log scale. The women's clean and jerk puts
six per cent of its mass on shapes at or above zero and crosses into that
region rarely, which is why it needed the longer chains. The Monte Carlo
standard error of the posterior median of the endpoint at the reference
bodyweight is about a tenth of a kilogram for the men's lifts and half a kilogram for the women's clean and jerk. Within the narrow region of the women's snatch it is $0.2$ kilograms, and the pooled median moves by less than $0.2$ kilograms when the weight of that region moves by one standard error. That of the one-in-a-thousand level
is under a tenth of a kilogram in all four combinations.
The replication package produces trace plots of both quantities. Of the
record-book fits, the one with the athlete factor for the men's clean and
jerk has chains that disagree on the threshold and the location
coefficient, with $\widehat{R}$ up to $1.47$, so its entries in Table~\ref{tab:ladder} of
the main text are indicative only. In the women's record-book fits the largest
values of $\widehat{R}$, up to $1.41$, again belong to the mixture
components, while the shape and the endpoint have $\widehat{R} \le 1.03$.
The threshold of the women's snatch fit with the athlete factor has an
effective sample size of $37$.

\begin{table}[t]
\centering
\caption{Convergence of the chains: acceptance rate, largest
$\widehat{R}$ over the parameters, smallest and median bulk effective
sample size over the parameters, and the bulk effective sample size of the
tail shape, the threshold and the endpoint at the reference bodyweight in
2026, with the $\widehat{R}$ of the endpoint, for the final model and for
the record-book fits without and with the athlete factor. Four
chains of $4{,}500$ draws for the final model ($16{,}500$ for the women's
lifts) and of $3{,}000$ for the record-book fits. For the women's snatch
the row for the third set of eight chains, which moves between the two
regions of the posterior, is followed by one row per region, with the
weight of the region in parentheses. The rows for the regions use the
draws of every chain in the region and compute $\widehat{R}$ over the
chains with at least $1{,}000$ draws there, cut to a common length.}
\label{tab:s-conv}
\resizebox{\textwidth}{!}{%
\begin{tabular}{lllrrrrrrrr}
\toprule
Sex & Lift & Fit & Acceptance & Max $\widehat{R}$ & Min ESS & Median ESS & $\xi_l$ & $u_l$ & Endpoint & $\widehat{R}$ endpoint \\
\midrule
Men & Snatch & Final model & 0.05 & 1.03 & 167 & 490 & 403 & 180 & 489 & 1.01 \\
Men & Snatch & Record book & 0.03 & 1.03 & 148 & 329 & 255 & 148 & 277 & 1.02 \\
Men & Snatch & Record book, athlete factor & 0.03 & 1.02 & 182 & 387 & 512 & 245 & 529 & 1.01 \\
Men & Clean and jerk & Final model & 0.04 & 1.03 & 112 & 383 & 252 & 117 & 417 & 1.03 \\
Men & Clean and jerk & Record book & 0.03 & 1.03 & 120 & 298 & 818 & 187 & 881 & 1.01 \\
Men & Clean and jerk & Record book, athlete factor & 0.02 & 1.47 & 8 & 82 & 159 & 23 & 143 & 1.05 \\
Women & Snatch & Final model, joint set & 0.01 & 1.24 & 25 & 98 & 55 & 35 & 157 & 1.06 \\
Women & Snatch & Final model, higher threshold (0.22) & 0.02 & 1.18 & 87 & 547 & 330 & 184 & 383 & 1.05 \\
Women & Snatch & Final model, lower threshold (0.78) & 0.02 & 1.33 & 103 & 310 & 417 & 242 & 553 & 1.05 \\
Women & Snatch & Record book & 0.02 & 1.29 & 11 & 89 & 128 & 82 & 159 & 1.02 \\
Women & Snatch & Record book, athlete factor & 0.02 & 1.24 & 15 & 111 & 317 & 37 & 348 & 1.00 \\
Women & Clean and jerk & Final model & 0.02 & 1.05 & 88 & 498 & 205 & 174 & 244 & 1.03 \\
Women & Clean and jerk & Record book & 0.02 & 1.41 & 9 & 32 & 183 & 31 & 256 & 1.01 \\
Women & Clean and jerk & Record book, athlete factor & 0.02 & 1.11 & 36 & 133 & 154 & 133 & 171 & 1.02 \\
\bottomrule
\end{tabular}}
\end{table}

\section{The whole archive}
\label{sec:s-archive}

Every analysis of the main text was repeated on the whole archive, the
$78{,}386$ athlete-day-lifts since 1998, with the same model, priors,
anchors and sampler settings. Table~\ref{tab:s-archive} compares the
final model on the two datasets and Table~\ref{tab:s-archive-ladder}
repeats Table~\ref{tab:ladder} of the main text on the whole archive.

The attempt and declaration parameters and the athlete correlation barely
change, and neither do the men's tails, which are bounded on both datasets
with shapes within $0.02$ of each other. The whole archive places the
men's ceilings at $110$ kilograms of bodyweight four and five kilograms
higher. The time trend of the location is much smaller on the whole
archive, $0.04$ to $0.06$ kilograms a year for men against $0.32$ and
$0.36$ on the current era and $0.35$ and $0.45$ for women against $0.60$
and $0.79$, so a single linear trend over the whole period does not
describe the rise within the current era. For women the whole archive
gives nearly the same shapes, $-0.22$ against $-0.23$ in the snatch and
$-0.21$ against $-0.18$ in the clean and jerk, with a wider interval for
the clean and jerk, a probability of a finite ceiling of $1.00$ in the
snatch and $0.88$ in the clean and jerk, and ceilings at $86$ kilograms of
$157$ and $196$ kilograms, against $143$ and $187$ on the current era. At
that bodyweight the heaviest women's lifts of all years, $138$ kilograms
in the snatch and $166$ in the clean and jerk, were made in 2009 and 2012,
and the model carries the tail they imply to 2026 through the trend in the
location. The women's clean and jerk tail on the whole archive depends more on the anchor. At $\tau = 0.01$ the posterior mode of its shape is
$-0.03$ and that of its endpoint $336$ kilograms, against $196$ and $198$
at $\tau = 0.02$ and $0.03$. At the lower-threshold mode of the women's
snatch the endpoint lies between $153$ and $157$ kilograms at the three
anchors.

On the whole archive the record book with the athlete-days exchangeable gives men's
ceilings of $212$ and $256$ kilograms, two and six kilograms above the
full model, and with the athlete factor $227$ and $276$, with the tail
still bounded. The ordering of the layers is the same as on the current era,
and the athlete factor again widens the record-book tail. The copula
gives $\rho = 0.96$ for men and $0.97$ for women, with $\psi = 0.15$ for
both. The chains of the final model on the whole archive mix as on the
current era for men, with $\widehat{R} \le 1.06$, and more slowly for
women. In the women's clean and jerk $\widehat{R}$ reaches $1.10$ and the
endpoint has an effective sample size of $83$. The women's snatch has the two regions of Section~\ref{sec:s-conv}. A joint set of four chains, run as there, gives $0.95$ of the mass, with a standard error of $0.02$, to the lower-threshold region, where the mixture components reach
$\widehat{R} = 1.23$ and the endpoint has an effective sample size of
$138$. The higher-threshold region holds $0.05$ of the mass. Only two
chains spent long stretches in it and they disagree within it, with
$\widehat{R}$ of $1.5$ for the shape, so its $939$ draws among the
$17{,}423$ pooled ones are the least reliable part of the whole-archive posterior. Without them the median ceiling of the women's snatch at $86$
kilograms moves from $157.2$ to $156.9$ kilograms and the probability of a
finite ceiling from $0.996$ to $1$.

\begin{table}[t]
\centering
\caption{The final model on the athlete-days since 2017 (the analysis of
the main text) and on the whole archive since 1998: number of
athlete-day-lifts, posterior median and $95$ per cent interval of the tail
shape, posterior probability of a finite endpoint, posterior medians of
the one-in-a-thousand level and of the endpoint at the reference
bodyweight in 2026, and the posterior medians of the athlete
correlation, the time trend of the location in kilograms a year and the
attempt-order shifts.}
\label{tab:s-archive}
\resizebox{\textwidth}{!}{%
\begin{tabular}{lllrrrrrrrrr}
\toprule
Sex & Lift & Data & $n$ & $\xi_l$ & $P(\xi_l < 0)$ & $q_{0.999}$ & Endpoint & $\rho_{a,l}$ & Trend & $\delta_2$ & $\delta_3$ \\
\midrule
Men & Snatch & since 2017 & 10{,}492 & $-$0.31 ($-$0.36, $-$0.25) & 1.00 & 199.3 & 206 & 0.906 & 0.32 & 0.66 & 0.75 \\
 &  & since 1998 & 22{,}272 & $-$0.33 ($-$0.37, $-$0.24) & 1.00 & 204.2 & 210 & 0.910 & 0.06 & 0.64 & 0.74 \\
Men & Clean and jerk & since 2017 & 10{,}179 & $-$0.40 ($-$0.44, $-$0.36) & 1.00 & 241.0 & 245 & 0.895 & 0.36 & 0.37 & 0.15 \\
 &  & since 1998 & 21{,}622 & $-$0.40 ($-$0.43, $-$0.35) & 1.00 & 245.6 & 250 & 0.897 & 0.04 & 0.33 & 0.09 \\
Women & Snatch & since 2017 & 9{,}695 & $-$0.23 ($-$0.29, 0.05) & 0.96 & 132.1 & 143 & 0.921 & 0.60 & 0.82 & 1.06 \\
 &  & since 1998 & 17{,}395 & $-$0.22 ($-$0.28, $-$0.05) & 1.00 & 142.0 & 157 & 0.904 & 0.35 & 0.83 & 1.10 \\
Women & Clean and jerk & since 2017 & 9{,}517 & $-$0.18 ($-$0.28, 0.14) & 0.94 & 167.8 & 187 & 0.929 & 0.79 & 0.43 & 0.24 \\
 &  & since 1998 & 17{,}097 & $-$0.21 ($-$0.42, 0.19) & 0.88 & 178.4 & 196 & 0.910 & 0.45 & 0.41 & 0.27 \\
\bottomrule
\end{tabular}}
\end{table}

\begin{table}[t]
\centering
\caption{Table~\ref{tab:ladder} of the main text on the whole archive: posterior median
and $95$ per cent interval of the tail shape, and the posterior median of
the endpoint at the reference bodyweight in 2026 with, in parentheses, the
posterior probability that it lies within ten per cent of the local record
of the current era. The last row uses no body, with the point estimate of
the endpoint.}
\label{tab:s-archive-ladder}
\resizebox{\textwidth}{!}{%
\begin{tabular}{llrrrr}
\toprule
 & & Men, snatch & Men, c\&j & Women, snatch & Women, c\&j \\
 & & (record 200) & (record 242) & (record 127) & (record 161) \\
\midrule
Record book & $\xi_l$ & $-$0.24 ($-$0.26, $-$0.20) & $-$0.22 ($-$0.25, $-$0.18) & $-$0.09 ($-$0.20, 0.03) & $-$0.30 ($-$0.34, $-$0.25) \\
 & ceiling & 212 (1.00) & 256 (1.00) & 174 (0.00) & 182 (0.00) \\
Record book, athlete factor & $\xi_l$ & $-$0.22 ($-$0.26, $-$0.15) & $-$0.20 ($-$0.25, $-$0.08) & $-$0.13 ($-$0.23, 0.03) & $-$0.22 ($-$0.27, $-$0.17) \\
 & ceiling & 227 (0.01) & 276 (0.01) & 175 (0.00) & 202 (0.00) \\
Full model & $\xi_l$ & $-$0.33 ($-$0.37, $-$0.24) & $-$0.40 ($-$0.43, $-$0.35) & $-$0.22 ($-$0.28, $-$0.05) & $-$0.21 ($-$0.42, 0.19) \\
 & ceiling & 210 (0.99) & 250 (1.00) & 157 (0.00) & 196 (0.00) \\
No body (check) & $\xi_l$ & $-$0.24 ($-$0.30, $-$0.19) & $-$0.23 ($-$0.28, $-$0.18) & $-$0.30 ($-$0.36, $-$0.23) & $-$0.33 ($-$0.41, $-$0.24) \\
 & ceiling & 209 & 253 & 146 & 179 \\
\bottomrule
\end{tabular}}
\end{table}

\section{The copula stage}
\label{sec:s-copula}

The joint likelihood of the paired athlete-days was evaluated on the grid
\begin{align*}
\rho &\in \{0.88, 0.90, 0.92, 0.93, 0.94, 0.95, 0.96, 0.97, 0.98\},\\
\psi &\in \{0, 0.05, 0.10, 0.15, 0.20, 0.25, 0.30, 0.40, 0.50\}
\end{align*}
at five posterior draws of the marginal parameters of each lift, with the
athlete factor switched off, and the posterior over the grid averaged
across the draws. For men the posterior mass sits on $\rho = 0.95$ and
$0.96$ with $\psi = 0.15$ or $0.10$, and for women on $\rho = 0.97$ with
$\psi = 0.10$ or $0.05$. The Gaussian copula ($\psi = 0$) has posterior
probability $5 \times 10^{-12}$ for men and $10^{-5}$ for women. The grid
is coarse by design,
since each evaluation integrates the two capabilities of $10{,}174$ or
$9{,}511$ athlete-days over the product of the two quadrature grids. The
return level of the men's total at $110$ kilograms of bodyweight is
$439.7$ kilograms $(438.1, 442.0)$ at a horizon of one in a thousand
athlete-days and $445.8$ $(442.7, 449.0)$ at one in ten thousand, against
$440.1$ and $446.1$ for the sum of the two marginal levels on the same
$40$ posterior draws. For women at
$86$ kilograms the figures are $299.6$ against $300.0$ and $310.4$ against
$310.7$. Computed draw by draw, the median deficit of the total at one in
ten thousand is $0.3$ kilograms $(-0.4, 0.8)$ for men and $0.4$
$(-1.6, 2.4)$ for women.

\section{Proof of Proposition 1}
\label{sec:s-proof}

\begin{proof}[Proof of Proposition~\ref{prop:ridge}]
Drop the superscript $l$, write primes for quantities under
$T_\Delta \theta_l$ and put $z' = e^{\Delta} z$. Since $\mu_l$ and
$\omega_l$ are multiplied by $e^{\Delta}$ and $G_l$ is unchanged,
$g_i'(z') = e^{-\Delta} g_i(z)$ and $G_i'(z') = G_i(z)$, so the athlete
term $\kappa$ in \eqref{eq:lik} is unchanged. The margin
$\log(z'/w_{ij})$ and the truncation point $\log\{z'/(w_{i,j-1} + 1)\}$
exceed their values at $z$ by $\Delta$, as do the margin modes, while the
margin scales are unchanged, so every factor $d_k$ and $D_k$ in
\eqref{eq:unitlik} is unchanged. Finally, $\eta_{ij}'(z') = \eta_{ij}(z) +
\Delta/\tau_l + \delta_j' - \delta_j$, which equals $\eta_{ij}(z)$ for
$j = 2, 3$ and $\eta_{i1}(z) + \Delta/\tau_l$ for $j = 1$. Changing the
variable of the inner integral in \eqref{eq:lik} from $z'$ to $z$, the
factor $e^{\Delta}$ in $dz'$ cancels the factor $e^{-\Delta}$ in $g_i'$.
\end{proof}

\end{document}